\documentclass[10pt]{article}
\usepackage{amsmath}
\usepackage{graphicx}
\usepackage{color}
\usepackage{orcidlink}
\usepackage{float}
\usepackage{hyperref}
\hypersetup{colorlinks=true, linkcolor=blue, citecolor=blue, urlcolor=blue}
\usepackage{caption}
\usepackage{subcaption}
\usepackage{setspace}
\usepackage{multirow}
\usepackage{multicol}
\usepackage{amssymb}
\RequirePackage[numbers,sort&compress]{natbib}
\begin{document}
\baselineskip=16pt

\begin{center}
\LARGE{Black holes in general relativity coupled with NEDs surrounded by PFDM: thermodynamics, epicyclic oscillations, QPOs, and shadow}
\end{center}

\vspace{0.2cm}

\begin{center}
{\bf Faizuddin Ahmed\orcidlink{0000-0003-2196-9622}}\footnote{\bf faizuddinahmed15@gmail.com}\\
{\it Department of Physics, The Assam Royal Global University, Guwahati, 781035, Assam, India}\\
\vspace{0.2cm}
{\bf Sardor Murodov\orcidlink{0000-0003-2360-4475}}\footnote{\bf s.murodov@newuu.uz}\\{\it New Uzbekistan University, Movarounnahr Str. 1, Tashkent 100000, Uzbekistan}\\{\it Institute of Fundamental and Applied Research, National Research University TIIAME, Kori Niyoziy 39, Tashkent 100000, Uzbekistan}\\
\vspace{0.2cm}
{\bf Bekzod Rahmatov\orcidlink{0009-0001-0394-650X}}\footnote{\bf be.rahmatov@newuu.uz}\\{\it University of Tashkent for Applied Sciences, Str. Gavhar 1, Tashkent 100149, Uzbekistan}\\{\it Tashkent State Technical University, Tashkent 100095, Uzbekistan}\\

\vspace{0.2cm}

\end{center}

\vspace{0.2cm}

\begin{abstract}
In this work, we investigate the thermodynamics and motion of neutral test particles around a regular black hole immersed in a perfect fluid dark matter environment. We begin by examining the horizon structure and key thermodynamic properties, with particular emphasis on quantities such as the Hawking temperature and the specific heat capacity. These aspects provide important insight into the stability and physical behavior of the black hole system. We then proceed to analyze the dynamics of neutral test particles using the Hamiltonian formalism, through which we derive the effective potential governing particle motion. Using the effective potential, we further study quasiperiodic oscillations by determining the associated epicyclic frequencies and comparing them with available observational data. Using the observed QPO data of XTE J1550-564, GRO J1655-40, GRS 1915+105, and M82 X-1, we perform a Markov Chain Monte Carlo analysis to constrain the black hole mass, the magnetic charge parameter, the PFDM parameter, and the characteristic orbital radius. Finally, we investigate the black hole shadow and demonstrate how various geometric parameters influence its optical appearance. This analysis highlights the potential observational signatures of such black holes and their surrounding dark matter environment.
\end{abstract}

\tableofcontents

\section{Introduction}\label{sec:1}

Black holes occupy a central place in gravitational physics because they probe the nonlinear and strong-field regime of general relativity in its most extreme form. At the same time, the classical theory predicts that physically relevant black hole spacetimes generically contain curvature singularities, where the spacetime description itself ceases to be predictive and the standard laws of physics can no longer be applied in a meaningful way \cite{Hawking1973,HawkingPenrose1996}. This tension between the success of general relativity in describing black holes and its breakdown at the singularity has long been regarded as one of the clearest indications that Einstein gravity must eventually be completed or modified in the ultraviolet. In the absence of a full and universally accepted theory of quantum gravity, effective nonsingular geometries provide a useful intermediate framework in which one may explore how short-distance modifications of the gravitational field could alter the structure and phenomenology of compact objects.

A particularly influential route toward singularity resolution is furnished by regular black hole models. The earliest example is the Bardeen spacetime, proposed as a black hole geometry with an event horizon but without a curvature singularity \cite{Bardeen1968}. A decisive conceptual advance came when Ay\'on-Beato and Garc\'ia showed that regular black hole geometries can arise as exact solutions of Einstein gravity coupled to nonlinear electrodynamics (NED), and later reinterpreted the Bardeen spacetime as the gravitational field of a nonlinear magnetic monopole \cite{AyonBeatoGarcia1998,AyonBeatoGarcia1999,AyonBeatoGarcia2000}. This observation established NED not merely as a technical device, but as a physically motivated matter sector capable of supporting black hole geometries in which the central singularity is replaced by a regular core. Subsequent developments led to broader classes of regular electrically charged and magnetically charged solutions, clarified the role of de Sitter-like interiors, and revealed both the usefulness and the subtleties of NED-based constructions \cite{Bronnikov2001,Dymnikova2004,Berej2006,Hayward2006}.

The significance of nonlinear electrodynamics in black hole physics extends well beyond the issue of regularity itself. As a nonlinear completion of Maxwell theory, NED naturally appears in finite-field-strength models, in effective descriptions incorporating quantum corrections, and in low-energy settings inspired by string theory \cite{BornInfeld1934,Gonzalez2009,Wang2019}. From this viewpoint, NED-supported black holes are attractive not only because they soften or remove singular behavior, but also because they provide controlled laboratories in which non-Maxwell electromagnetic effects can backreact on spacetime geometry and influence observable quantities. In recent years, these models have therefore been investigated from multiple complementary directions, including thermodynamic stability, phase transitions, state-space geometry, quasinormal spectra, and rotating generalizations \cite{BalartVagenas2014,Tharanath2015,Gan2016,Wei2018,Toshmatov2017}. Such studies have made it increasingly clear that the physical content of a black hole solution depends not only on the metric itself, but also on the matter sector that supports it.

For astrophysical applications, however, the geometry of a compact object cannot be studied in isolation from its environment. Real black holes are immersed in external media such as accreting matter, electromagnetic fields, plasma, and very likely dark matter. Since a large body of observational evidence indicates that dark matter dominates the matter content of galactic halos, it is natural to ask how a dark-matter background modifies the structure and observational appearance of black holes. Among the phenomenological models used in this context, the perfect fluid dark matter (PFDM) framework has proved especially useful because it generates analytically tractable modifications of the metric while capturing some essential properties of dark halo distributions \cite{Rahaman2010}. In this approach, the spacetime typically acquires an additional logarithmic contribution governed by a dark-matter parameter, leading to potentially measurable changes in circular motion, horizon structure, lensing, and optical observables.

Black holes surrounded by PFDM have accordingly attracted growing attention in recent years. Their shadows and deflection properties have been analyzed in both static and rotating settings, and their thermodynamic structure and phase behavior have also been studied in detail \cite{Hou2018,Haroon2019,Xu2019,Hendi2020,Ahmed2026,Bouzenada2026,Sakalli2026,Silva2026,Badawi2026}. On the dynamical side, PFDM backgrounds have been shown to affect timelike geodesics, the location of stable circular orbits, and the associated epicyclic frequencies around both singular and regular black hole spacetimes \cite{Das2021,Rayimbaev2021,Zhang2021,Narzilloev2020}. These results are important because they indicate that environmental effects may mimic, mask, or amplify deviations produced by the intrinsic parameters of the black hole itself. In particular, once nonlinear electrodynamics and dark matter are both present, one expects a nontrivial interplay between short-distance regularization and large-scale matter distribution, and this interplay can only be understood through a systematic combined analysis.

A natural language for such an analysis is provided by the motion of test particles and photons. Timelike geodesics encode the structure of the effective potential, the existence and stability of circular orbits, and the position of the innermost stable circular orbit (ISCO), all of which are directly related to accretion physics and strong-gravity phenomenology. Null geodesics, on the other hand, determine the photon sphere, light bending, lensing properties, and the apparent black hole shadow. For regular black holes, these features have already been shown to differ in important ways from their Schwarzschild and Reissner--Nordstr\"om counterparts, particularly in the near-horizon and inner regions of the spacetime \cite{StuchlikSchee2015,Toshmatov2015QNM,Tang2023,Sun2023}. When a PFDM background is added, both the timelike and null sectors are modified further, making geodesic diagnostics an especially sensitive probe of the geometry.

Another crucial observable is furnished by quasi-periodic oscillations (QPOs) in X-ray binaries and related accreting compact systems. In many models, the observed QPO frequencies are associated with orbital motion and small oscillations around circular geodesics, and are therefore governed by the fundamental frequencies of the background spacetime. This makes QPOs a particularly sharp probe of strong-field gravity, since even modest deformations of the metric can shift the orbital, radial epicyclic, and vertical epicyclic frequencies in ways that are in principle detectable. Over the years, relativistic precession models, resonance models, and related approaches have been used to constrain the parameters of black holes and black-hole mimickers from timing data \cite{Kolos2015, Tursunov2016, Stuchlik2016, Stuchlik2021, Vrba2021, StellaVietri1998,Rezzolla2004,Sramkova2015,Kolos2017,Jiang2021,Wang2022,Liu2023,AlBadawi2026,Bouzenada2026,Ahmad2026,Orhan2026}. More recently, similar ideas have also been extended to black holes embedded in nonvacuum dark-sector backgrounds, further strengthening the case for combining geodesic analysis with timing phenomenology \cite{Rahmatov2024CJPhys,Ahmed2026,Fathi2026}.

Recent studies have shown that QPO-based Bayesian inference can serve as a powerful tool for testing strong-gravity spacetimes and constraining the physical parameters of compact objects. In particular, Murodov and collaborators applied QPO analyses to several non-standard black hole backgrounds and demonstrated that timing data can place meaningful bounds on spacetime parameters beyond the standard Schwarzschild or Kerr descriptions. For example, in their study of magnetized Bocharova–Bronnikov–Melnikov–Bekenstein black holes, they showed that QPO signals and circular-motion properties provide an effective framework for probing deviations from general relativity \cite{Murodov2025EPJC}. A related analysis of regular Ayón-Beato–García black holes further emphasized that QPO observations can be used to test regular black hole geometries and to constrain the parameters governing particle dynamics in such backgrounds \cite{RahmatovMurodovRayimbaev2025}. More recently, this line of research was extended to the Johannsen–Psaltis metric, where QPO studies were combined with accretion modeling to test the strong gravitational field regime in a more general phenomenological setting \cite{DonmezMurodovRayimbaev2026}. Motivated by these developments, in the present work we employ an MCMC framework to infer the black hole mass and the additional spacetime parameters of the regular black hole immersed in PFDM from the observed twin-peak QPO data.

At the same time, horizon-scale imaging has opened a complementary and exceptionally powerful observational window. The Event Horizon Telescope images of M87* and Sgr A* have transformed black hole shadow studies from a purely theoretical subject into a precision tool for testing the geometry of compact objects in the strong-field regime \cite{EHTL1,EHTL4,EHTL6,EHTL12,EHTL16,EHTL17}. Since NED effects can modify photon propagation and since dark-matter environments may alter the size and deformation of the shadow, the optical appearance of a black hole provides an independent channel through which one may test such models. This is particularly relevant for regular black holes, whose shadow and lensing observables have been shown to depend sensitively on the underlying matter source and on whether photon trajectories follow the background geometry or an effective geometry induced by nonlinear electrodynamics \cite{Tang2023,Sun2023}. A coherent treatment of particle motion, QPO-related frequencies, and shadow observables is therefore essential if one wishes to assess the astrophysical viability of any NED-supported black hole embedded in a dark-matter background.

These considerations provide the main motivation for the present work. Closely related studies have recently examined black hole spacetimes in general relativity coupled to nonlinear electrodynamics and surrounded by PFDM, as well as the corresponding dynamics of test particles, thermodynamic behavior, and optical signatures \cite{RayimbaevRahmatovZahid2025,Rahmatov2025EPJC,Rahmatov2025PDUlensing}. Building on this line of research, here we investigate in a unified framework the geometry and phenomenology of a black hole solution supported by nonlinear electrodynamics in the presence of perfect fluid dark matter. Our goal is to determine how the simultaneous presence of a nonlinear electromagnetic source and a dark-matter environment affects the horizon structure, thermodynamic quantities, timelike geodesics, circular orbits and ISCO properties, fundamental frequencies relevant for QPO models, and the null-geodesic structure responsible for the black hole shadow. In this way, the present analysis is intended to clarify which features originate from the intrinsic NED sector, which arise from the external PFDM background, and which observational signatures may help disentangle the two.

The paper is organized as follows. Section~\ref{sec2} presents the regular black hole spacetime in the presence of PFDM and outlines its main properties. In Sec.~\ref{sec3}, we examine the horizon structure and thermodynamic behavior of the solution. Section~\ref{sec4} is devoted to the dynamics of neutral test particles, where we study the effective potential, radial force, circular orbits, ISCO behavior, particle trajectories, and the associated fundamental frequencies. In Sec.~\ref{qpo}, we discuss the QPO phenomenology in the framework of the relativistic precession model. Section~\ref{sec6} contains the MCMC analysis and the resulting constraints on the model parameters from observational data. In Sec.~\ref{Sec:shadow}, we study null geodesics and the black hole shadow. Finally, Sec.~\ref{sec8} presents our conclusions and a discussion of the physical significance of the results.

\section{Regular Black hole Immersed in PFDM}\label{sec2}

In this section, we study a regular black hole solution immersed in a PFDM background. We analyze its thermodynamic properties as well as the epicyclic frequencies of neutral test particles, highlighting the effects of the magnetic charge and the PFDM parameter on the dynamics of the system. 

Let us consider Einstein gravity coupled to a non-linear electromagnetic field in the presence of the perfect fluid dark matter, described by the following equations \cite{Simovic2024,Cadoni2023}:
\begin{align}
G_{\mu\nu}&= 2 \left[ \frac{\partial \mathcal{L}(\mathcal{F})}{\partial \mathcal{F}} 
F_{\mu\lambda} F_{\nu}{}^{\lambda} - g_{\mu\nu} \mathcal{L}(\mathcal{F}) \right]+T^{\rm PFDM}_{\mu\nu}, \label{eq1} \\
\nabla_{\mu} \left( \frac{\partial \mathcal{L}(\mathcal{F})}{\partial \mathcal{F}}F^{\mu\nu} \right) &= 0, \label{eq2} \\
\nabla_{\mu} \left( {}^{\ast}F^{\mu\nu} \right) &= 0, \label{eq3}
\end{align}

where $G_{\mu\nu}$ is the Einstein tensor, $T^{\rm PFDM}_{\mu\nu}$ is the energy-momentum tensor of PFDM, $F_{\mu\nu}$ is the Maxwell tensor with ${}^{\ast}F^{\mu\nu}$ is its dual. We consider the nonlinear electrodynamics Lagrangian density $\mathcal{L}(\mathcal{F})$, where $\mathcal{F} = \frac{1}{4} F^{\mu\nu} F_{\mu\nu}$, given by \cite{SG2024}
\begin{align} 
{\mathcal{L}}({\mathcal{F}}) = \frac{24{\mathcal{F}}}{s \left(2^\frac{3}{4}+2{\mathcal{F}}^{\frac{1}{4}}\sqrt{q}\right)^4}, \label{eq4}
\end{align}
where the parameter s is related to the magnetic charge and the mass of the black hole via  $s=q/(2M)$. Further, we consider the Maxwell field tensor as follows
\begin{align} 
F_{\mu \nu } = 2\delta ^{\theta }_{[\mu }\delta ^{\phi }_{\nu ]}q(r)\sin {\theta }. 
\end{align}
Consequently, the magnetic field strength is given by
\begin{align} 
F_{23}= 2 q \sin {\theta }, \quad \mbox{and} \quad {\mathcal{F}} = \frac{q^2}{2r^4}.\label{eq5} 
\end{align}
By substituting the value of $\mathcal{F}$ from Eq. (\ref{eq5}) into Eq. (\ref{eq4}), we obtain
\begin{align} 
{\mathcal{L}}(r) =\frac{3Mq}{(r+q)^4}.\label{eq5b}
\end{align}

Moreover, the energy-momentum of perfect fluid dark matter is given by \cite{MHL2012}
\begin{equation}
T^{\text{DM}}_{\,\,\mu\nu} = \mathrm{diag}\left(-\mathcal{E}_{DM},\, P_{r\,DM},\, P_{\theta\,DM},\, P_{\phi\,DM}\right),\label{eq6}
\end{equation}
where the components are given by
\begin{equation}
\mathcal{E}_{DM} = -P_{r\,DM}= -\frac{\lambda}{8\pi r^3},\label{eq7}
\end{equation}
with 
\begin{equation}
P_{\theta\,DM}= -\frac{\lambda}{16\pi r^3}=P_{\phi\,DM}.\label{eq8}
\end{equation}
Here $\lambda$ is a constant that takes real values, called PFDM parameter and the equation of state is $p_{\theta\,DM}=(\delta-1) \mathcal{E}_{DM}$ with $\delta=3/2$.

To obtain the regular black hole solution immersed in a perfect fluid dark matter, we assume the static and spherically symmetric ansatz as given by
\begin{equation}
    ds^2 = -f(r)\,dt^2 + \dfrac{dr^2}{f(r)} + r^2 (d\theta ^2 + \sin ^2{\theta }\,d\varphi ^2),\label{metric}
\end{equation}
The function \(f(r)\) is to be determined by solving the field Eq. (\ref{eq1}). 

Finally, solving Einstein equation, Eq. (\ref{eq2}), using Eqs.  (\ref{eq6}) and (\ref{metric}), we obtain the solution:
\begin{align}
     f(r) = 1-\frac{2 M r^2}{(r+q)^3}+\frac{\lambda}{r}\ln\!\frac{r}{|\lambda|},\label{function}
\end{align}
Here $M$ represent mass of the black hole, $q$ is the magnetic charge parameter. Noted that this metric is a very special case of the general space-time reported in \cite{CPC2025}, where one can set $\nu=1, \mu=3$. In the limit $q=0$, corresponding to the absence of magnetic charge, the considered black hole simplifies to the Schwrazschild metric surrounded by perfect fluid dark matter, reported in \cite{MHL2012}

\section{Horizon Structure and Thermodynamics}\label{sec3}

In this section, we study the horizon structure of the considered black hole and analyze the effects of the model parameters in comparison with the standard Schwarzschild black hole.

The event horizon represents the boundary beyond which nothing, not even light, can escape. It is determined by the condition that the lapse function vanishes, namely $f(r_h)=0$. Thus, the horizon equation takes the form
\begin{equation}
1-\frac{2 M r_h^2}{(r_h+q)^3}+\frac{\lambda}{r_h}\ln\!\frac{r_h}{|\lambda|}=0.
\label{bb1}
\end{equation}

\begin{figure}[ht!]
    \centering
    \includegraphics[width=0.45\linewidth]{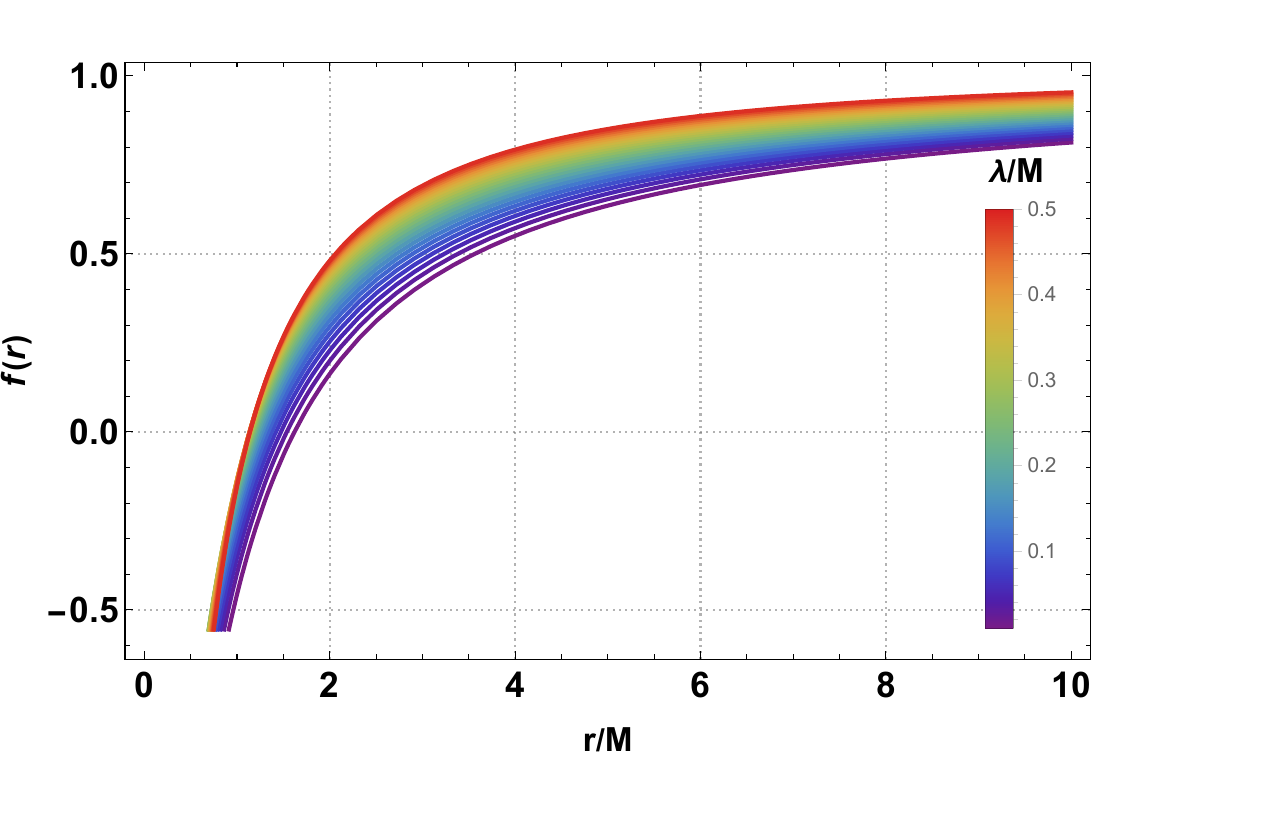}
    \includegraphics[width=0.45\linewidth]{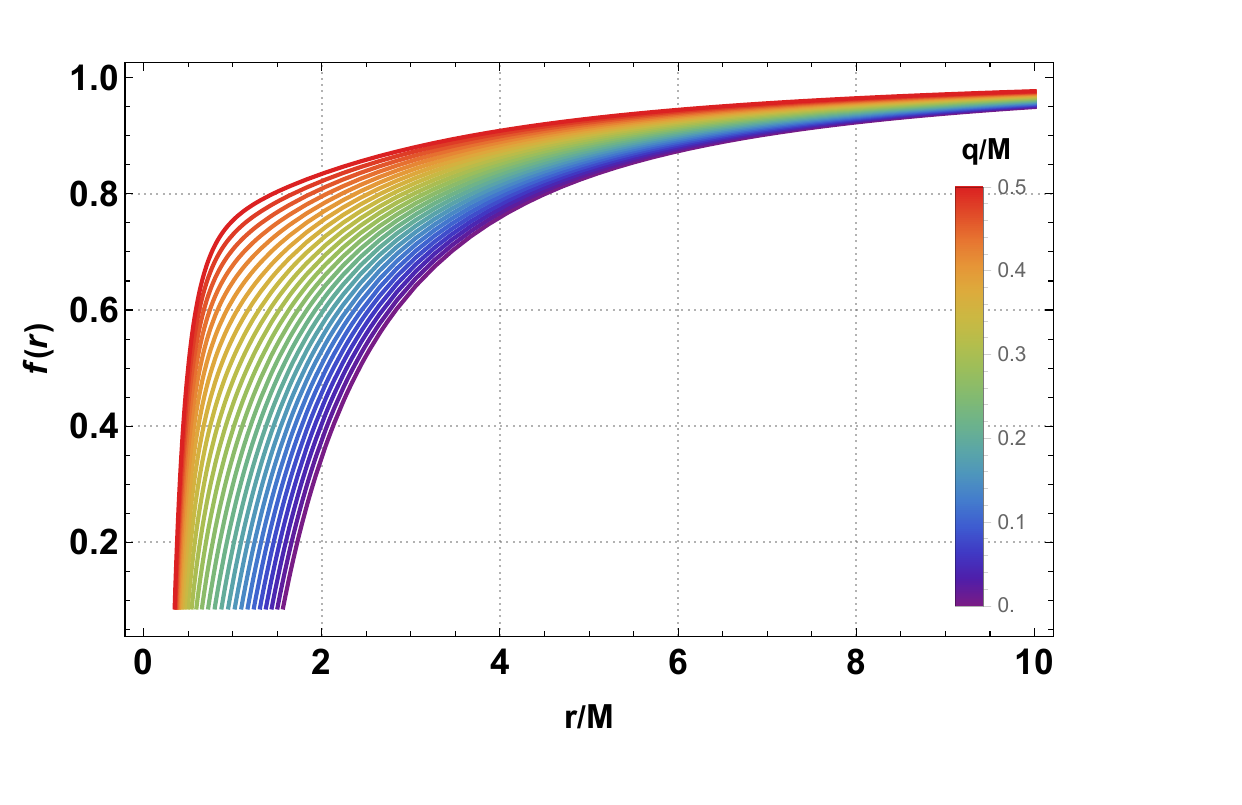}\\
    (i) $q/M=0.1$ \hspace{6cm} (ii) $\lambda/M=0.5$
    \caption{Behavior of the metric function for different values of $\lambda$ and $q$.}
    \label{fig:1}
\end{figure}

In Figure~\ref{fig:1}, we illustrate the lapse function as a function of the dimensionless radial coordinate for different values of the PFDM parameter $\lambda$ and the magnetic charge $q$. In both panels, we observe that as these parameters increase, the metric function $f(r)$ shifts upward, leading to a decrease in the event horizon radius. 

\begin{table}[ht!]
\centering
\begin{tabular}{|c|c|c|c|c|c|c|c|c|c|}
\hline
$\lambda/M$ 
& -0.9 & -0.8 & -0.7 & -0.6 & -0.5 & -0.4 & -0.3 & -0.2 & -0.1 \\
\hline
$r_h/M$ 
& 1.62403 & 1.65612 & 1.67415 & 1.67854 & 1.66928 & 1.64579 & 1.60666 & 1.54881 & 1.46484 \\
\hline
$\lambda/M$ 
& 0.1 & 0.2 & 0.3 & 0.4 & 0.5 & 0.6 & 0.7 & 0.8 & 0.9 \\
\hline
$r_h/M$ 
& 1.18646 & 1.13768 & 1.11573 & 1.10921 & 1.11291 & 1.12395 & 1.14053 & 1.16148 & 1.18597 \\
\hline
\end{tabular}
\caption{Event horizon radius for various values of $\lambda$, with fixed $q/M=0.1$.}
\label{tab:1}
\end{table}

\begin{table}[ht!]
\centering
\begin{tabular}{|c|c|c|c|c|c|c|c|c|c|}
\hline
$\lambda/M$ 
& -0.9 & -0.8 & -0.7 & -0.6 & -0.5 & -0.4 & -0.3 & -0.2 & -0.1 \\
\hline
$r_h/M$ 
& 1.49764 & 1.54126 & 1.56780 & 1.57847 & 1.57373 & 1.55326 & 1.51576 & 1.45805 & 1.37219 \\
\hline
$\lambda/M$ 
& 0.1 & 0.2 & 0.3 & 0.4 & 0.5 & 0.6 & 0.7 & 0.8 & 0.9 \\
\hline
$r_h/M$ 
& 1.08021 & 1.03183 & 1.01249 & 1.00965 & 1.01755 & 1.03299 & 1.05402 & 1.07933 & 1.10805 \\
\hline
\end{tabular}
\caption{Event horizon radius for various values of $\lambda$, with fixed $q/M=0.2$.}
\label{tab:2}
\end{table}

In tables \ref{tab:1}--\ref{tab:2}, we presented the numerical results for the event horizon $r_h$ by varying the PFDM parameter $\lambda$ for two values of the magnetic charge $q=0.1$ and $0.2$.

From Eq.~(\ref{bb1}), the ADM mass can be written as
\begin{equation}
M=\frac{(r_h+q)^3}{2r_h^2}\left[1+\frac{\lambda}{r_h}\ln\!\frac{r_h}{|\lambda|}\right].
\label{bb2}
\end{equation}

Since the space-time is asymptotically flat, with $\lim_{r\to\infty}f(r)=1$, the Hawking temperature is related to the surface gravity $\kappa=f'(r_h)/2$ through \cite{Hawking1974,Hawking1975,Bardeen1973}
\begin{equation}
T_H=\frac{\kappa}{2\pi}=\frac{f'(r_h)}{4\pi}.
\end{equation}
Using the metric function, we obtain
\begin{equation}
T_H=\frac{1}{4\pi r_h}\left[
\frac{2 M r_h^2 (r_h-2q)}{(r_h+q)^4}
+\frac{\lambda}{r_h}\left(1-\ln\!\frac{r_h}{|\lambda|}\right)
\right].
\label{bb3}
\end{equation}
In the limit $q=0$, the Hawking temperature reduces to
\begin{equation}
T_H=\frac{1}{4\pi r_h}\left[
\frac{2M}{r_h}+\frac{\lambda}{r_h}\left(1-\ln\!\frac{r_h}{|\lambda|}\right)
\right].
\label{bb4}
\end{equation}

\begin{figure}[ht!]
    \centering
    \includegraphics[width=0.5\linewidth]{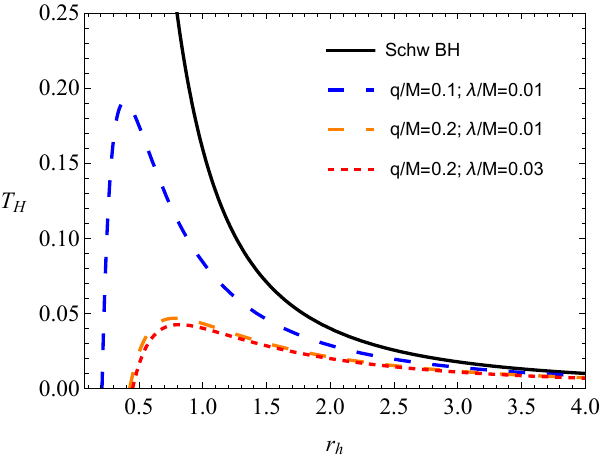}
    \caption{Hawking temperature $T_H$ versus the event horizon radius $r_h$ for the Schwarzschild black hole and for different values of $q/M$ and $\lambda/M$. The Schwarzschild case decreases monotonically, while the modified solutions show a peak and lower temperatures.}
    \label{fig:th}
\end{figure}

Figure~\ref{fig:th} shows that the Hawking temperature of the Schwarzschild black hole decreases monotonically with increasing event horizon radius, which is the standard thermodynamic behavior. In contrast, the modified black hole solutions exhibit a non-monotonic temperature profile: $T_H$ starts from a very small value, rises to a finite maximum, and then decreases gradually as $r_h$ increases. This behavior indicates that the parameters $q/M$ and $\lambda/M$ significantly modify the thermal structure of the system.

It is also evident that increasing $q/M$ lowers the peak value of the Hawking temperature and shifts the corresponding curve downward. A similar suppressing effect is observed when $\lambda/M$ increases. Therefore, both parameters reduce the thermal activity of the black hole and soften the evaporation process relative to the Schwarzschild case. At large $r_h$, all curves tend to approach each other, showing that the effects of these parameters become less important for larger black holes and that the solution gradually approaches the Schwarzschild limit. The vanishing temperature in the small-radius region may also indicate the possibility of an extremal configuration or a black hole remnant.

For a thermodynamic system obeying the first law of black hole thermodynamics, the entropy is given by one-quarter of the Bekenstein--Hawking horizon area \cite{Bekenstein1973,Hawking1975},
\begin{equation}
S=\frac{A}{4}=\pi r_h^2.
\label{bb5}
\end{equation}

The modified first law of thermodynamics can be written as
\begin{equation}
dM=T_H\,dS+\Psi_q\,dq+\Phi_{\lambda}\,d\lambda,
\label{bb6}
\end{equation}
where $\Phi_{\lambda}$ is the thermodynamic potential associated with the PFDM parameter $\lambda$, and $\Psi_q$ is the potential conjugate to the magnetic charge parameter $q$. These quantities are given by
\begin{align}
\Phi_{\lambda}
&=\left(\frac{\partial M}{\partial \lambda}\right)_{S,q}
=\frac{(r_h+q)^3}{2r_h^3}
\left(\ln\!\frac{r_h}{|\lambda|}-1\right),
\label{bb7}\\
\Psi_q
&=\left(\frac{\partial M}{\partial q}\right)_{S,\lambda}
=\frac{3(r_h+q)^2}{2r_h^2}
\left(1+\frac{\lambda}{r_h}\ln\!\frac{r_h}{|\lambda|}\right).
\label{bb8}
\end{align}

Using the above relations, one finds that the thermodynamic quantities satisfy the modified Smarr relation
\begin{equation}
M=2T_HS+q\Psi_q+\lambda\Phi_{\lambda}.
\label{bb9}
\end{equation}

Next, we determine the heat capacity of the thermodynamic system, which is given by
\begin{align}
\mathrm{C}
=\frac{2\pi r_h^2\,\mathcal{C}}
{
\mathcal{D}-\mathcal{C}
}.
\label{bb10}
\end{align}
where
\begin{align}
    \mathcal{D}&=\frac{3q r_h}{(r_h+q)^2}
\left(1+\frac{\lambda}{r_h}\ln\frac{r_h}{|\lambda|}\right)
+\frac{\lambda}{r_h}
\left[
\frac{r_h-2q}{r_h+q}\left(1-\ln\frac{r_h}{|\lambda|}\right)
-\left(2-\ln\frac{r_h}{|\lambda|}\right)
\right]\nonumber\\
    \mathcal{C}&=
\frac{r_h-2q}{r_h+q}
\left(1+\frac{\lambda}{r_h}\ln\!\frac{r_h}{|\lambda|}\right)
+\frac{\lambda}{r_h}\left(1-\ln\!\frac{r_h}{|\lambda|}\right).
\end{align}

In the limit $q=0$, the heat capacity simplifies to
\begin{equation}
C=-2\pi r_h^2\left(\frac{1+\lambda/r_h}{1+2\lambda/r_h}\right),
\label{bb11}
\end{equation}
which further reduces to the Schwarzschild result when $\lambda=0$.

\begin{figure}
    \centering
    \includegraphics[width=0.5\linewidth]{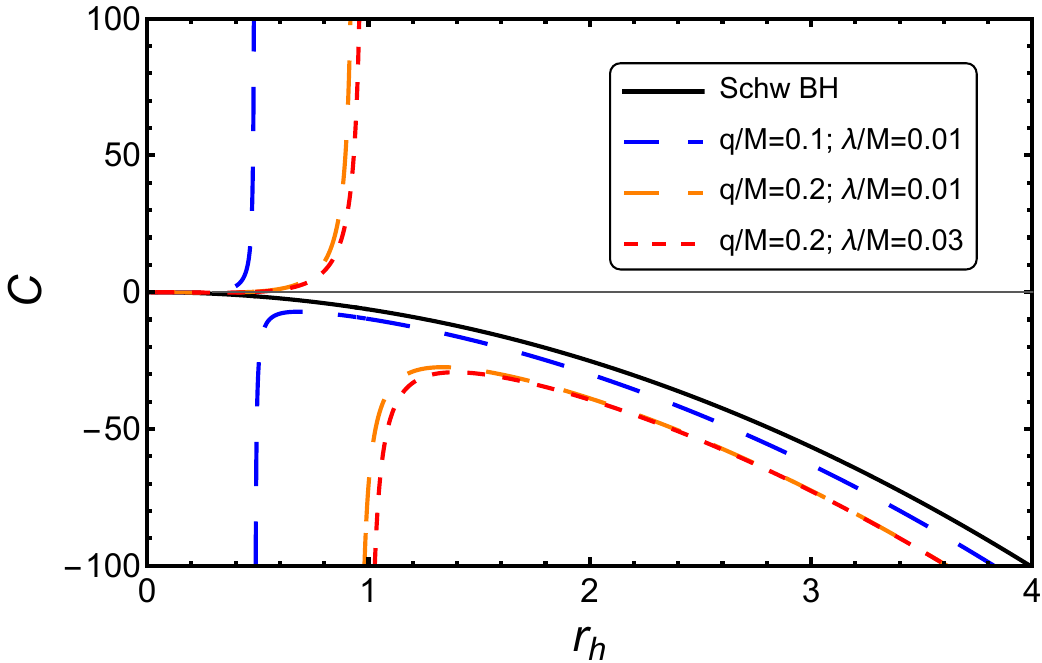}
   \caption{Heat capacity $C$ as a function of the event horizon radius $r_h$ for the Schwarzschild black hole and for the modified black hole solutions with different values of $q/M$ and $\lambda/M$. The Schwarzschild case remains negative over the whole range, whereas the modified solutions exhibit divergences at critical radii and a sign change of $C$, indicating a thermodynamic phase transition. The horizontal line $C=0$ separates the locally stable and unstable branches.}
    \label{fig:С}
\end{figure}

Figure \ref{fig:С} shows the thermodynamic behavior of the black hole through the heat capacity $C$, obtained from the thermodynamic relations of the model. In the Schwarzschild case, the heat capacity is always negative and decreases monotonically with increasing horizon radius, which is the standard signature of thermodynamic instability. In contrast, the modified black hole solutions display a qualitatively different behavior: each curve develops a divergence at a certain critical radius, where the heat capacity changes sign. Such divergences indicate second-order phase transition points separating different thermodynamic branches.
For the modified cases, the small-radius branch with $C>0$ corresponds to a locally thermodynamically stable configuration, while the branch with $C<0$ corresponds to an unstable one. Thus, unlike the Schwarzschild black hole, the presence of the parameters $q/M$ and $\lambda/M$ creates a stable thermodynamic region in the small-horizon domain. This behavior is consistent with the non-monotonic Hawking-temperature profile discussed earlier: since $C=dM/dT_H$, the extrema of $T_H$ are directly associated with the divergence of the heat capacity. 
It is also seen that increasing the deformation parameters shifts the divergence point toward larger values of $r_h$ and moves the negative branch downward. This means that stronger deviations from the Schwarzschild geometry enhance the modification of the thermal response and change the location of the transition between stable and unstable phases. At sufficiently large $r_h$, all curves remain negative, showing that large black holes recover the usual unstable Schwarzschild-like thermodynamic behavior. 

Finally, to study the global thermodynamic stability of the system, we consider the Gibbs free energy, defined as
\begin{equation}
G=M-T_HS.
\label{bb12}
\end{equation}
Substituting Eqs.~(\ref{bb2})--(\ref{bb5}), one can analyze the dependence of $G$ on the horizon radius and the model parameters in order to identify thermodynamically favored configurations.
\section{Motion of Neutral Test Particles}\label{sec4}

In this section, we study the motion of neutral test particles around the considered black hole, deriving the effective potential and analyze it. Through the Hamiltonian formalism, we derive the effective potential that governs the particles dynamics and analyze innermost stable circular orbits. Moreover, we determine the epicyclic frequencies using this potential shown how various geometric parameters alter these frequencies.

The motion of neutral test particle of mass $m$ is described by the Hamiltonian of the system as,
\begin{equation}
    H=\frac{1}{2} g^{\mu\nu} p_{\mu} p_{\nu}+\frac{1}{2} m^2,\label{cc1}
\end{equation}
where $g_{\mu\nu}$ is the metric tensor and $p_{\mu}$ is the four-momenta related with four-velocity by $p^{\mu}=m u^{\mu}$, while $u^{\mu}=dx^{\mu}/d\tau$ with $\tau$ being proper time.

Now, for the Hamiltonian equations of motion, we consider the form
\begin{equation}
\frac{dx^\alpha}{d\zeta} \equiv m u^\alpha = \frac{\partial H}{\partial p_\alpha}, \quad
\frac{dp_\alpha}{d\zeta} = - \frac{\partial H}{\partial x^\alpha}, 
\label{cc2}
\end{equation}
where the affine parameter can be assumed as $\zeta = \tau / \mu$.

Essentially, the black hole geometry has two constants of motion: (i) the energy $E$, and the specific angular momentum $L_z$. In our case at hand, we find these are 
\begin{align}
\frac{p_t}{m} &=-f(r) \frac{dt}{d\tau}=-\mathcal{E}, \label{cc3} \\
\frac{p_\phi}{m} &= r^2 \sin^2 \theta \frac{d\phi}{d\tau}=\mathcal{L}_z, \label{cc4}\\
\frac{p_{\theta}}{m}&=r^2 \frac{d\theta}{d\tau}=\mathcal{L}_{\theta},\label{cc5}
\end{align}
where $\mathcal{E}=E/m$ and $\mathcal{L}_z=L_z/m$ represent the energy and angular momentum per unit mass of test particles, respectively. The total angular momentum is then given by $\mathcal{L}^2=\mathcal{L}^2_z/\sin^2 \theta+\mathcal{L}^2_{\theta}$.

Hence, the components of the 4-velocity $u^\mu$ are
\begin{align}
\frac{dt}{d\tau} &= \frac{\mathcal{E}}{1 -\frac{2 M r^2}{(r+q)^3}+\frac{\lambda}{r}\ln\!\frac{r}{\lambda}}, \label{cc6} \\
\frac{d\phi}{d\tau} &= \frac{\mathcal{L}_z}{r^2 \sin^2 \theta}, \label{cc7} \\
\frac{d\theta}{d\tau}&=\frac{1}{r^2}\sqrt{\mathcal{L}^2-\mathcal{L}^2_z/\sin^2 \theta},\label{cc8}\\
\frac{dr}{d\tau}&=\sqrt{\mathcal{E}^2-\left(1+ \frac{\mathcal{L}^2}{r^2} \right)\left(1 -\frac{2 M r^2}{(r+q)^3}+\frac{\lambda}{r}\ln\!\frac{r}{\lambda}\right)}, \label{cc9}
\end{align}

In our case, the Hamiltonian takes the form
\begin{equation}
H = \frac{1}{2} \left(1 -\frac{2 M r^2}{(r+q)^3}+\frac{\lambda}{r}\ln\!\frac{r}{\lambda}\right) p_r^2
  + \frac{m^2}{2 f(r)}\left[U_{\rm eff}(r)- E^2\right].
\label{cc10}
\end{equation}

The effective potential of the Hamiltonian system is then given by
\begin{equation}
U_{\rm eff}(r)=\left(1+ \frac{\mathcal{L}^2}{r^2} \right)\left(1 -\frac{2 M r^2}{(r+q)^3}+\frac{\lambda}{r}\ln\!\frac{r}{\lambda}\right).\label{cc11}
\end{equation}
From the above analysis, we observe that $\mathcal{L}_z=\mathcal{L}$ when $\theta=\pi/2$.

\begin{figure}
    \centering
    \includegraphics[width=0.5\linewidth]{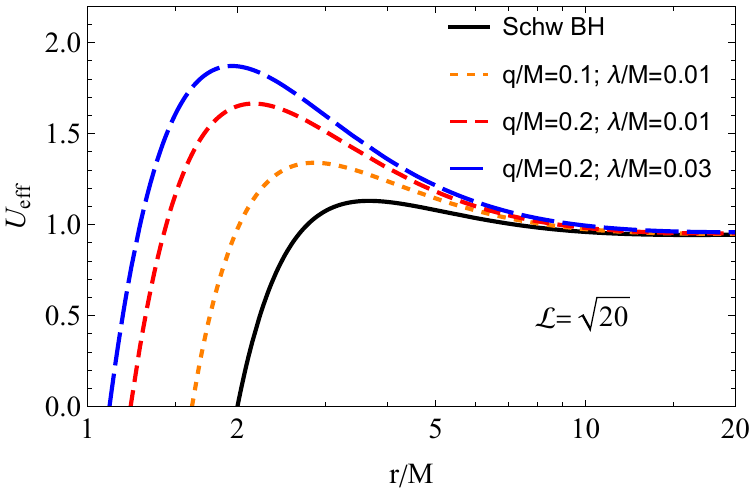}
    \caption{Effective potential $U_{\rm eff}$ as a function of $r/M$ for different values of $q$ and $\lambda$ with fixed angular momentum $\mathcal{L}=20$. The plot shows that the magnetic charge and PFDM parameters noticeably affect the orbital structure of test particles.}
    \label{ueff}
\end{figure}

Fig. \ref{ueff} illustrates how the effective potential changes with the radial coordinate for different choices of the magnetic charge parameter $q$ and the PFDM parameter $\lambda$. Compared with the Schwarzschild case, increasing $q$ and $\lambda$ modifies the shape of the potential curve and shifts its minimum toward larger radii. This behavior indicates that stable circular orbits are formed farther from the black hole. At the same time, the reduction in the depth of the potential well suggests that test particles become less strongly bound in the gravitational field. Physically, this means that both the magnetic charge and the surrounding perfect fluid dark matter environment play an important role in altering the particle dynamics around the compact object. As a consequence, the innermost stable circular orbit can move outward, the corresponding orbital frequencies may decrease, and the inner edge of the accretion disk may be displaced to larger distances. These effects can lead to observable changes in astrophysical phenomena such as disk radiation and quasi-periodic oscillations.

\subsection{Effective Radial Force}

The effective force acting on a test particle indicates whether it is (i) attracted toward or (ii) repelled away from the black hole. In essence, a black hole can exert gravitational forces that either draw in or push away particles. With this in mind, we carefully analyze the motion of the test particles and compute the effective force using Eq. (\ref{cc11}) as follows:
\begin{equation}
    \mathcal{F}=-\frac{1}{2}\,\frac{\partial^2U_{\rm eff}}{\partial r^2}=\frac{M r (2q - r)}{(r+q)^4} - \frac{\lambda}{2r^2} \left( 1 - \ln \frac{r}{\lambda} \right)+\frac{\mathcal{L}^2}{r^3}\,\left(1 - \frac{\lambda}{2r} - \frac{3 M r^3}{(r+q)^4} + \frac{3 \lambda}{2r} \ln \frac{r}{\lambda}\right).\label{force} 
\end{equation}

\begin{figure}
    \centering
    \includegraphics[width=0.5\linewidth]{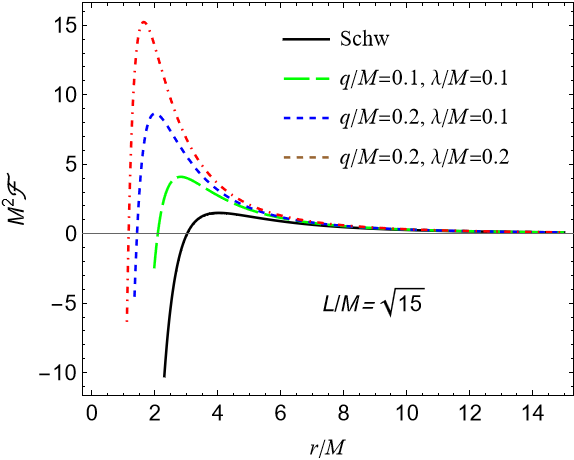}
    \caption{Effective radial force $M^{2}\mathcal{F}$ as a function of $r/M$ for different values of $q/M$ and $\lambda/M$, with fixed angular momentum $L/M=\sqrt{15}$.}
    \label{fig:force}
\end{figure}

Form the above expression, we observe that geometric parameters $\lambda$ and $q$ influence the force experienced by the test particles moving around the black hole. Moreover, the conserved angular momentum $\mathcal{L}$ alter this force.

In the limit $q=0$, corresponding to the absence of the magnetic charge, the expression of force simplifies as
\begin{equation}
    \mathcal{F}=-\frac{M}{r^2}- \frac{\lambda}{2r^2} \left( 1 - \ln \frac{r}{\lambda} \right)+\frac{\mathcal{L}^2}{r^3}\,\left(1 - \frac{\lambda}{2r} - \frac{3 M}{r} + \frac{3 \lambda}{2r} \ln \frac{r}{\lambda}\right).\label{force-2} 
\end{equation}
Equation (\ref{force-2}) is the force experienced by the test particles in the Schwrazschild black hole surrounded by PFDM.

Fig. \ref{fig:force} shows that the effective radial force is strongly affected by the magnetic charge parameter $q$ and the PFDM parameter $\lambda$. Compared with the Schwarzschild case, the modified black hole solutions develop a higher peak in the small-$r$ region, indicating a stronger influence of the spacetime geometry on particle motion near the black hole. It is also seen that increasing $q$ and $\lambda$ enhances the magnitude of the force and shifts its maximum toward smaller radii. At larger distances, however, all curves gradually approach zero, showing that the effects of these parameters become weaker far from the black hole.

\subsection{Circular Motions: ISCO analysis}

For circular motion of test particles, we have the condition $U_{\rm eff}(r)=\mathcal{E}^2$ and $dU_{\rm eff}/dr=0$. Simplification of these conditions results
\begin{align}
    \mathcal{L}^2_{\rm }&=\frac{r^3\,f'(r)}{2 f(r)-r\,f'(r)}=r^2\,\frac{ - \frac{M r^2 (2q - r)}{(r+q)^4} + \frac{\lambda}{2r} \left( 1 - \ln \frac{r}{\lambda} \right)}{1 - \frac{\lambda}{2r} - \frac{3 M r^3}{(r+q)^4} + \frac{3 \lambda}{2r} \ln \frac{r}{\lambda}}.\label{cc12}\\
    \mathcal{E}^2_{\rm }&=\frac{2\,f^2(r)}{2 f(r)-r\,f'(r)}=\frac{\left(1 -\frac{2 M r^2}{(r+q)^3}+\frac{\lambda}{r}\ln\!\frac{r}{\lambda}\right)^2}{1 - \frac{\lambda}{2r} - \frac{3 M r^3}{(r+q)^4} + \frac{3 \lambda}{2r} \ln \frac{r}{\lambda}}.\label{cc13}
\end{align}

\begin{figure}
    \centering
    \includegraphics[width=0.34\linewidth]{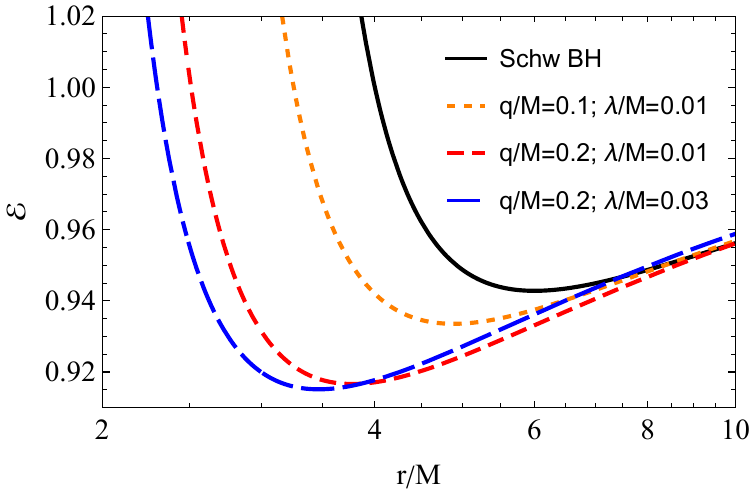}
    \includegraphics[width=0.32\linewidth]{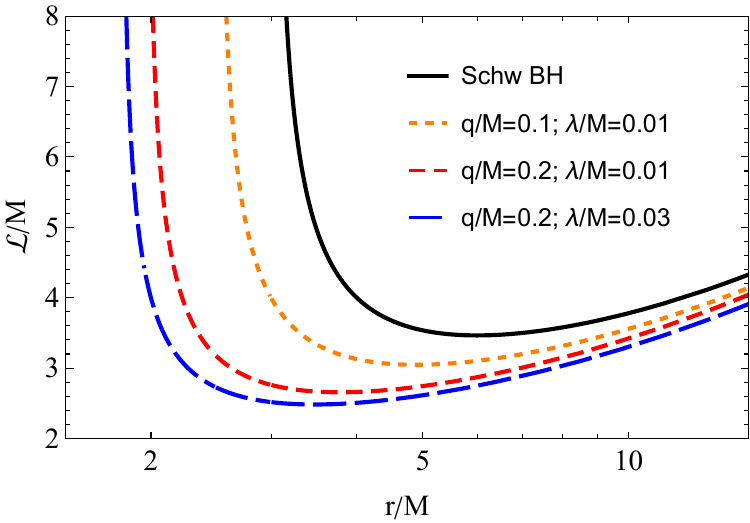}
    \includegraphics[width=0.28\linewidth]{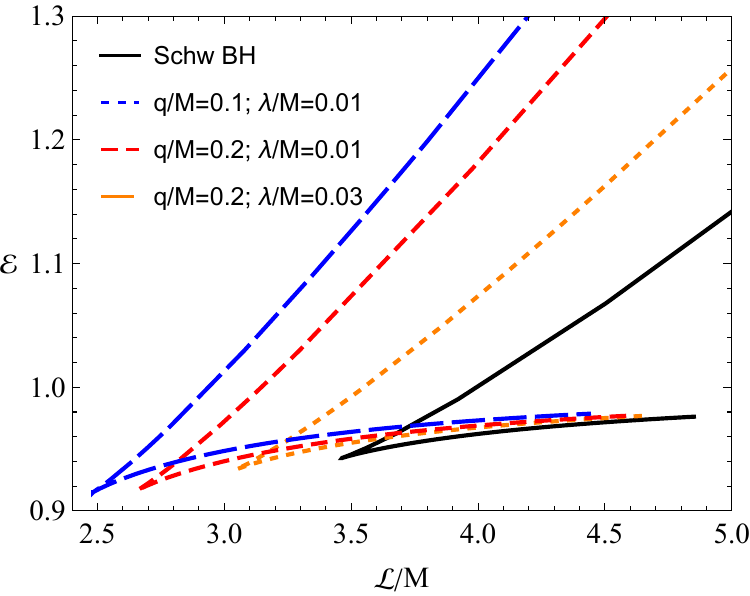}
  \caption{Energy $\mathcal{E}$ and angular momentum $\mathcal{L}/M$ for circular motion around the black hole for different values of $q$ and $\lambda$. The left and middle panels show their radial profiles, while the right panel presents the $\mathcal{E}$--$\mathcal{L}/M$ relation. The black solid curve corresponds to the Schwarzschild case.}
    \label{eell}
\end{figure}

Figure \ref{eell} illustrates the combined effect of the magnetic charge parameter $q$ and the PFDM parameter $\lambda$ on the energetics and angular momentum of circular orbits around the black hole. In general, the presence of these parameters modifies the orbital structure significantly compared with the Schwarzschild case. The shifts in the minima of the curves indicate changes in the location and properties of stable circular orbits, while the $\mathcal{E}$--$\mathcal{L}$ relation shows how the energy cost of orbital motion is altered in the modified spacetime. These changes are important for understanding the motion of test particles, the position of the innermost stable circular orbit, and possible astrophysical signatures in accretion and oscillation phenomena.
In the left panel, the specific energy $\mathcal{E}$ is plotted as a function of the radial coordinate $r/M$. The minima of the curves correspond to stable circular orbits. It can be seen that the magnetic charge and PFDM parameters shift the minimum energy to smaller radii and slightly reduce its value relative to the Schwarzschild case. This behavior suggests that the modified geometry allows bound circular orbits to exist deeper in the gravitational field.
In the middle panel, the angular momentum $\mathcal{L}/M$ is shown as a function of $r/M$. The minima again mark the transition to the innermost stable circular orbit. The decrease of the minimum angular momentum for larger $q$ and $\lambda$ indicates that stable motion can be maintained with lower angular momentum in the modified spacetime. This reflects the strong impact of the magnetic charge and dark matter environment on orbital dynamics.
In the right panel, the parametric dependence of the specific energy on angular momentum is presented. The modified cases deviate clearly from the Schwarzschild curve, showing that the balance between orbital energy and angular momentum is sensitive to both $q$ and $\lambda$. This result implies that the efficiency of accretion processes and the dynamical behavior of orbiting matter may differ noticeably from the standard Schwarzschild black hole case.

For marginally stable orbits, we have the additional condition $d^2U_{\rm eff}/dr^2 \geq 0$ including the aforementioned conditions. For innermost stable circular orbits, we find the following relations:
\begin{equation}
    f(r)\,f''(r)-2 (f'(r))^2+3\,f(r)\,f'(r)/r=0.\label{cc14}
\end{equation}

\begin{table}[ht!]
\centering
\begin{tabular}{|c|c|c|c|c|c|c|c|}
\hline
$\lambda/M (\downarrow) \backslash q/M (\rightarrow)$ & 0.00 & 0.05 & 0.10 & 0.15 & 0.20 & 0.25 & 0.30 \\
\hline
-0.5 & 8.3741 & 8.0302 & 7.6787 & 7.3184 & 6.9475 & 6.5637 & 6.1635 \\
-0.4 & 8.1113 & 7.7613 & 7.4036 & 7.0371 & 6.6600 & 6.2699 & 5.8635 \\
-0.3 & 7.7918 & 7.4319 & 7.0640 & 6.6868 & 6.2983 & 5.8960 & 5.4759 \\
-0.2 & 7.3956 & 7.0203 & 6.6361 & 6.2413 & 5.8333 & 5.4089 & 4.9630 \\
-0.1 & 6.8775 & 6.4771 & 6.0653 & 5.6396 & 5.1961 & 4.7290 & --- \\
\hline
\end{tabular}
\caption{Variation of the ISCO radius $r_{\text{ISCO}}/M$ with $q/M$ and $\lambda/M$.}
\label{tab:3}
\end{table}

\begin{figure}[ht!]
    \centering
    \includegraphics[width=0.65\linewidth]{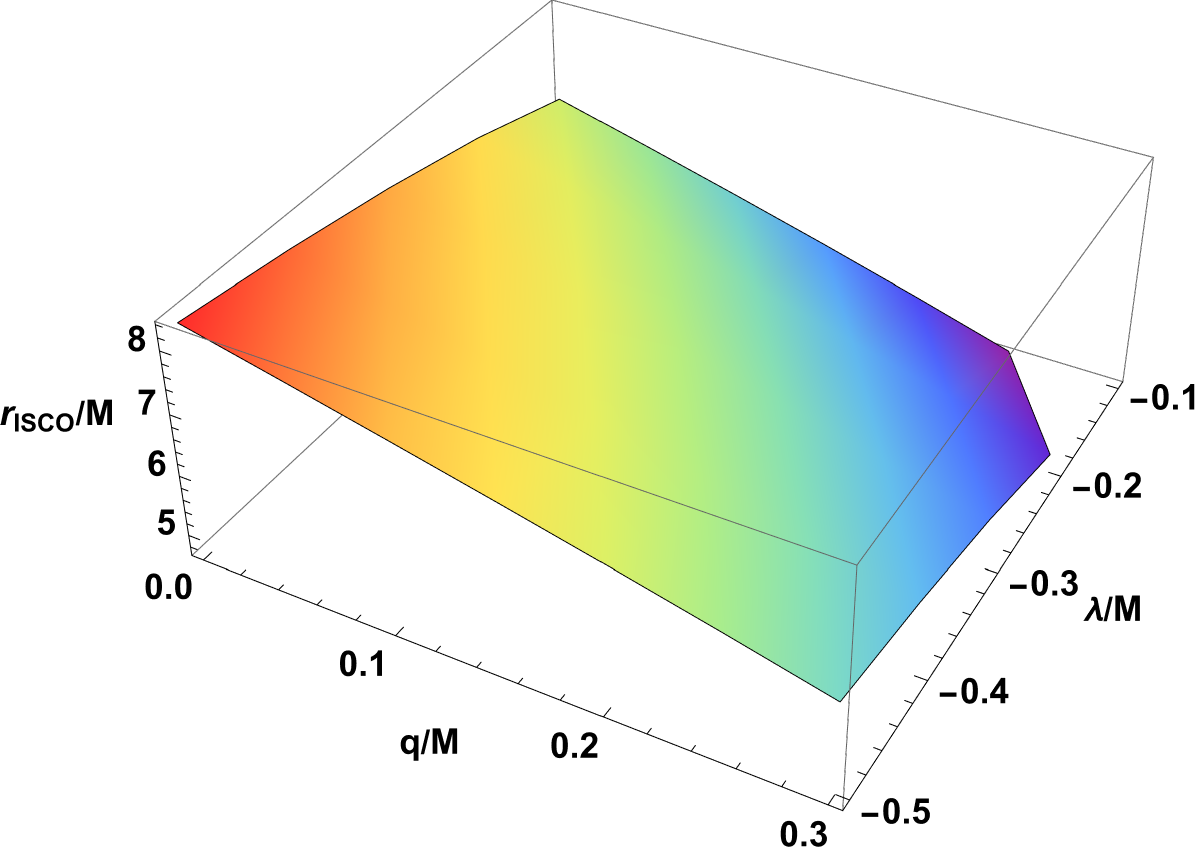}
    \caption{Three-dimensional plot of ISCO radius for various $q$ and $\lambda$.}
    \label{fig:2}
\end{figure}

Substituting the metric metric function $f(r)$ into the Eq. (\ref{cc14}) and solving for $r$ will give us the ISCO radii. Noted that the exact analytical solution of this polynomial equation in $r$ is quite impossible due to the presence of the logarithmic function. Hence, by choosing suitable values of parameters $q$ and $\lambda$, we determine numerical values of the ISCO radii.

In Table \ref{tab:3}, we presented numerical results for the ISCO radius by varying both the PFDM parameter $\lambda$ and the magnetic charge $q$. We observe that for a particular value of $\lambda$, an increasing $q$ shrinks the ISCO size. Similar patter can be seen for a fixed $q$ and increasing $\lambda$. Collectively, both geometric parameters influence the size of the particles stable orbit. 

In Figure~\ref{fig:2}, we present a graphical representation of the ISCO radius as a function of the PFDM parameter $\lambda$ and the magnetic charge $q$.

\subsection{Test Particle Trajectories}

The motion of test particles around a black hole is determined by the underlying spacetime geometry, resulting in a rich variety of possible trajectories such as bound orbits, scattering paths, or eventual capture. In the strong-field region near the event horizon, relativistic effects become dominant, leading to pronounced orbital precession and significant deviations from classical motion \cite{Chandrasekhar1983}. 

In the equatorial plane defined by $\theta=\pi/2$, the equation of orbit for the test particles using Eqs. (\ref{cc7}) and (\ref{cc9}) ia given by
\begin{equation}
    \left(\frac{1}{r^2}\frac{dr}{d\phi}\right)^2=\frac{\mathcal{E}^2}{\mathcal{L}^2}-\left(\frac{1}{\mathcal{L}^2}+\frac{1}{r^2}\right)\,f(r).\label{ee1}
\end{equation}
Transforming to anew variable via $u(\phi)=\frac{1}{r(\phi)}$ and after simplification result:
\begin{equation}
\left(\frac{du}{d\phi}\right)^2=\frac{\mathcal{E}^2}{\mathcal{L}^2}-\left(\frac{1}{\mathcal{L}^2}+u^2\right)\,\left(1-\frac{2 M u}{(1+q u)^3}-\lambda u\,\ln\!(|\lambda|\,u)\right).\label{ee2}
\end{equation}

Differentiating both sides w. ro to $\phi$ and after simplification results
\begin{align}
    \frac{d^2u}{d\phi^2}
=
-\frac{1}{2}\Bigg[
2u\left(\frac{1}{\mathcal{L}^2}+u^2\right)\left(1-\frac{2Mu}{(1+qu)^3}-\lambda u\ln(|\lambda|u)\right)
+\left(\frac{1}{\mathcal{L}^2}+u^2\right)
\left(
-\frac{2M(1-2qu)}{(1+qu)^4}
-\lambda(\ln(|\lambda|u)+1)
\right)
\Bigg].\label{ee3}
\end{align}

\begin{figure}[ht!]
    \centering
    \includegraphics[width=0.45\linewidth]{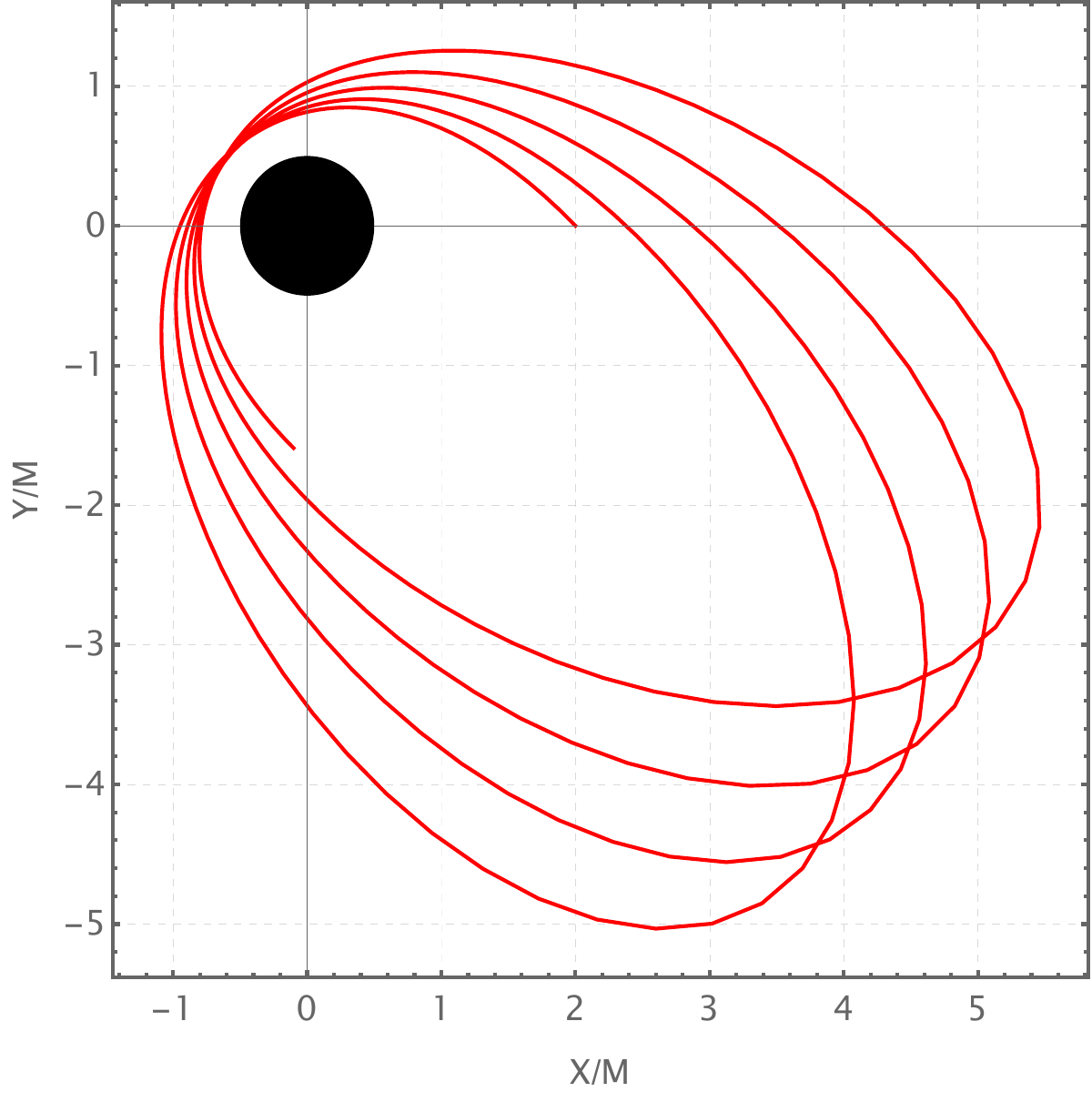}\qquad
    \includegraphics[width=0.45\linewidth]{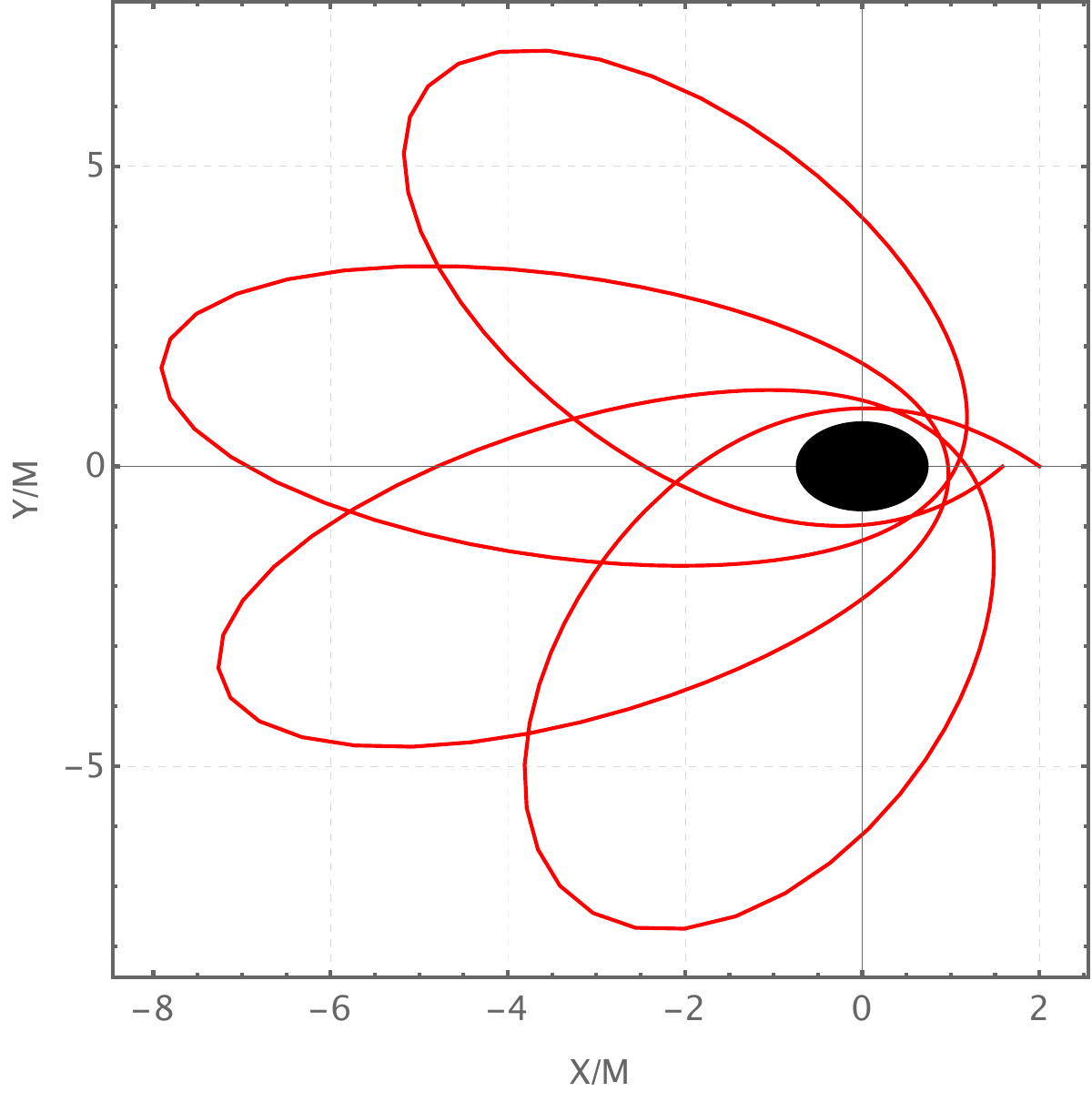}\\
    (i) $\lambda=-0.1,\,q=0.5$ \hspace{4cm} (ii) $\lambda=-0.1,\,q=0.6$ 
    \caption{Particle trajectory in the $(X,Y)$ plane corresponding to $\mathcal{L}=1$, and $M=1$.}
    \label{fig:trajectory}
\end{figure}

The above second-order non-linear differential equation (\ref{ee3}) represents the test particles trajectory in the considered black hole gravitational field. We observe that the geometric parameters $\lambda$ and $q$ together with the conserved angular momentum $\mathcal{L}$ alter the particles trajectory in the field.

In Figures~\ref{fig:trajectory}, we illustrate the particle trajectories in the $(X,Y)$ plane for different values of the PFDM parameter $\lambda$ and the magnetic charge $q$, while keeping $\mathcal{L}$ fixed. We observe that, for a fixed value of $\lambda=-0.1$, even a small variation in $q$ leads to a significant change in the particle trajectory.

\subsection{Fundamental Frequencies}

We now determine the fundamental frequencies associated with circular motion of neutral test particles in the considered spacetime. In order to avoid notational ambiguity, we distinguish between the frequencies measured with respect to the coordinate time \(t\), denoted by \(\Omega_i\), and those measured with respect to the proper time \(\tau\), denoted by \(\omega_i\).

The azimuthal, or Keplerian, frequency measured by a static observer at infinity is given by
\begin{equation}
\Omega_{\phi}
=
\frac{d\phi}{dt}\Big|_{r=r_c}
=
\sqrt{\frac{f'(r)}{2r}}
=
\sqrt{
-\frac{M(2q-r)}{(r+q)^4}
+
\frac{\lambda}{2r^3}
\left(
1-\ln\frac{r}{\lambda}
\right)
}.
\label{dd1}
\end{equation}
Because of the spherical symmetry of the metric, the vertical epicyclic frequency coincides with the azimuthal one, namely
\begin{equation}
\Omega_{\theta}=\Omega_{\phi}.
\end{equation}

The radial epicyclic frequency, also measured with respect to the coordinate time, is obtained from the second derivative of the effective potential and can be written as
\begin{equation}
\Omega_r
=
\sqrt{
\frac{1}{2}
\left[
f(r)f''(r)-2\bigl(f'(r)\bigr)^2+\frac{3f(r)f'(r)}{r}
\right]
}.
\label{dd2}
\end{equation}

The corresponding locally measured frequencies are related to the above quantities through
\begin{equation}
\omega_i=u^t\Omega_i,
\qquad
u^t=\frac{dt}{d\tau}=\left(f-\frac{r f'}{2}\right)^{-1/2},
\end{equation}
where \(i=\phi,\theta,r\). Therefore,
\begin{equation}
\omega_{\phi}=u^t\Omega_{\phi},
\qquad
\omega_{\theta}=u^t\Omega_{\theta},
\qquad
\omega_r=u^t\Omega_r.
\end{equation}

Hence, \(\Omega_{\phi}\) and \(\Omega_r\) represent the orbital and radial oscillation frequencies seen by a distant observer, while \(\omega_{\phi}\), \(\omega_{\theta}\), and \(\omega_r\) correspond to the locally measured frequencies. In the following, the plots are constructed using \(\Omega_{\phi}\) and \(\Omega_r\).

\begin{figure}
\includegraphics[scale=0.61]{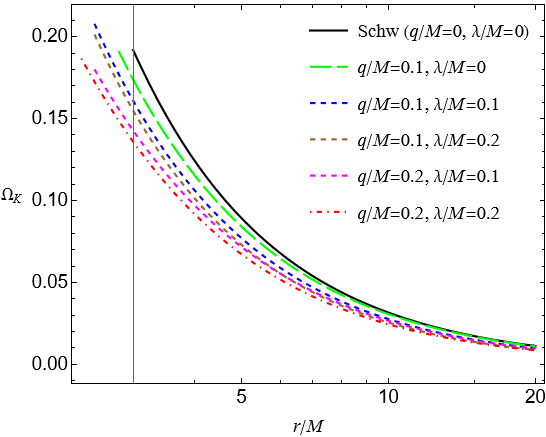}
\includegraphics[scale=0.6]{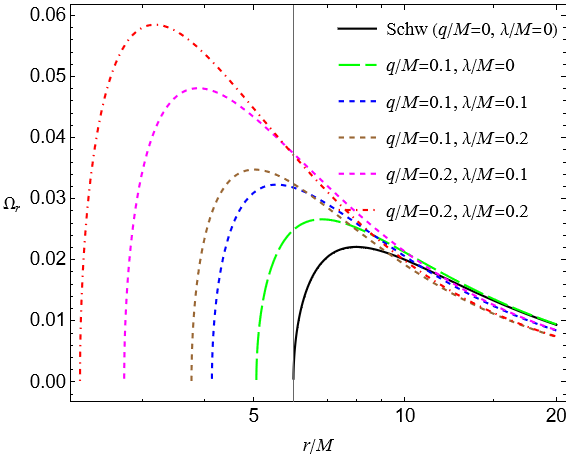}
\caption{\label{fig:Kep_rad}Left panel: radial epicyclic frequency $\Omega_r$ as a function of $r/M$ for different values of the magnetic charge parameter $q$ and the PFDM parameter $\lambda$. Right panel: correlation between the upper and lower QPO frequencies, $\nu_U$ and $\nu_L$, in the RP model for the same set of parameters. The Schwarzschild case is shown by the black solid curve, while the colored dashed curves correspond to the regular black hole immersed in PFDM.}
\end{figure}
Figure~\ref{fig:Kep_rad} illustrates how the magnetic charge parameter $q$ and the PFDM parameter $\lambda$ modify both the radial epicyclic motion and the QPO frequency relation. In the left panel, the radial epicyclic frequency $\Omega_r$ vanishes at small radii, reaches a maximum, and then gradually decreases at larger distances. Compared with the Schwarzschild case, increasing $q$ and $\lambda$ shifts the curves toward smaller radii and enhances the peak value of $\Omega_r$. This behavior indicates that the non-vacuum structure of the spacetime has a strong influence on the radial oscillation properties of test particles in the inner accretion region. In the right panel, the RP frequency diagram shows that the modified spacetime systematically moves the $\nu_U$--$\nu_L$ trajectories upward relative to the Schwarzschild limit. Therefore, for a fixed lower frequency, the model predicts a larger upper frequency when either $q$ or $\lambda$ increases. This result demonstrates that both the magnetic charge and the surrounding perfect fluid dark matter environment leave clear imprints on the QPO spectrum, making these frequencies useful probes of the underlying black hole geometry.

\section{Quasi-Periodic Oscillations  \label{qpo}}
QPOs are among the most important timing signatures observed in accreting compact objects. In black hole X-ray binaries and ultraluminous X-ray sources, they appear as narrow features in the power density spectrum and are generally interpreted as manifestations of characteristic dynamical frequencies in the inner accretion flow. Since these frequencies originate in the strong-gravity region close to the compact object, they provide a useful observational probe of the underlying spacetime geometry and of the physical parameters of the source \cite{vanderKlis2006,Belloni2012}.

Among the proposed theoretical interpretations of high-frequency QPOs, the relativistic precession (RP) model has proven to be particularly useful \cite{Stella1998,Stella1999}. In this framework, the observed twin-peak QPOs are associated with combinations of the fundamental frequencies of nearly circular geodesic motion. More precisely, the upper frequency is identified with the azimuthal orbital frequency, while the lower frequency is related to the periastron precession frequency. Therefore, the model predicts
\begin{equation}
\nu_U = \nu_{\phi}, \qquad
\nu_L = \nu_{\phi} - \nu_r ,
\end{equation}
where $\nu_{\phi}$ is the orbital frequency and $\nu_r$ is the radial epicyclic frequency. This prescription makes the RP model especially suitable for studying how deviations from the standard Schwarzschild geometry affect observable timing properties.

In the present work, we apply the RP model to the regular black hole immersed in perfect fluid dark matter. Since both the magnetic charge parameter $q$ and the PFDM parameter $\lambda$ modify the metric function, they also alter the orbital and radial epicyclic frequencies. As a consequence, the theoretical relation between the lower and upper QPO frequencies is shifted relative to the Schwarzschild case. This behavior suggests that QPO observations can be used to constrain not only the black hole mass, but also the additional parameters introduced by the nonlinear electromagnetic sector and the surrounding dark matter distribution.

To confront the model with observational data, we consider four well-known sources for which twin-peak QPO frequencies have been reported: the microquasars XTE~J1550--564, GRO~J1655--40, and GRS~1915+105, together with the ultraluminous X-ray source M82~X-1 \cite{Remillard2002,Strohmayer2001,Morgan1997,Pasham2014}. These systems span a broad mass range, from stellar-mass black holes to intermediate-mass black hole candidates, and therefore provide a useful testing ground for the model. The observed pairs of frequencies are employed in the following section to statistically constrain the physical parameters of the spacetime.

\begin{figure}
    \centering
    \includegraphics[width=0.49\linewidth]{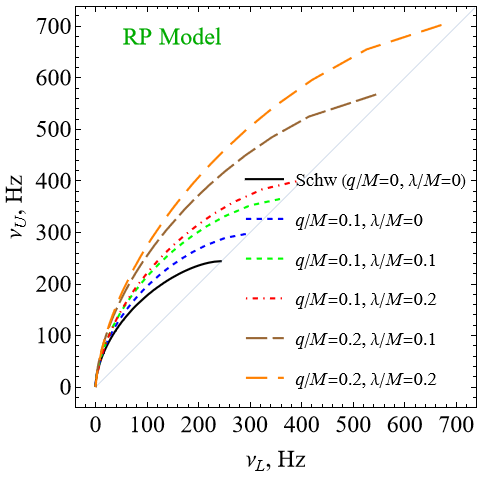}
   \caption{$\nu_U$ versus $\nu_L$ in the RP model for different $q$ and $\lambda$. The black curve shows the Schwarzschild case, while the dashed curves represent the PFDM black hole. Larger $q$ and $\lambda$ shift the curves upward.} \label{modes}
\end{figure}

Figure - \ref{modes} presents the correlation between the upper and lower QPO frequencies in the framework of the RP model for several values of the magnetic charge parameter $q$ and the PFDM parameter $\lambda$. The Schwarzschild case is shown by the black solid curve, whereas the remaining curves correspond to the regular black hole surrounded by perfect fluid dark matter. A clear deviation from the Schwarzschild behavior is observed once the parameters $q$ and $\lambda$ are introduced.

For all cases, the upper frequency $\nu_U$ increases monotonically with the lower frequency $\nu_L$, which is consistent with the general behavior expected in the relativistic precession scenario. However, the relation is strongly nonlinear, and the theoretical tracks lie above the reference line $\nu_U=\nu_L$, reflecting the fact that in the RP model the upper frequency is always larger than the lower one. More importantly, the inclusion of the parameters $q$ and $\lambda$ systematically shifts the curves upward. This means that, for a given value of $\nu_L$, the model predicts a larger $\nu_U$ compared with the Schwarzschild spacetime.

It is also evident that the magnetic charge parameter produces a significant enhancement of the frequency relation. For example, the cases with $q/M=0.2$ are located noticeably above those with $q/M=0.1$, indicating that stronger magnetic charge leads to higher predicted upper frequencies. The PFDM parameter $\lambda$ acts in a similar way: for fixed $q$, increasing $\lambda$ moves the corresponding trajectory further upward. Therefore, both parameters contribute constructively to modifying the QPO spectrum.

From a physical point of view, this behavior shows that the magnetic structure of the regular black hole and the surrounding dark matter distribution both affect the orbital and epicyclic motion of test particles in the inner accretion region. Since the RP frequencies are directly built from these fundamental frequencies, even moderate changes in the spacetime parameters produce visible shifts in the $\nu_U$--$\nu_L$ diagram. This sensitivity makes the observed QPO pairs a useful tool for constraining the parameters of the model and for distinguishing the present spacetime from the standard Schwarzschild geometry.

\section{MCMC analysis}\label{sec6}

High-time-resolution X-ray observations make it possible to study the dynamics of matter in the immediate vicinity of compact objects through QPOs, which appear as narrow peaks in the power density spectra of accreting sources \cite{Gendreau2016NICER,vanderKlis1989}. Since these oscillatory features are linked to the characteristic frequencies of particle motion in the strong-gravity region, they provide an efficient observational tool for constraining both the spacetime geometry and the physical parameters of black holes.

In the present work, we focus on four systems for which twin high-frequency QPOs have been reported: XTE~J1550--564, GRO~J1655--40, GRS~1915+105, and M82~X-1 \cite{Orosz2011,Strohmayer2001,Miller2015,Pasham2014}. The first three objects are stellar-mass black hole candidates in microquasars, whereas M82~X-1 is an ultraluminous X-ray source widely discussed as a possible intermediate-mass black hole candidate \cite{Pasham2014,Stuchlik2015M82,Casella2008,Dewangan2006}. The observed upper and lower QPO frequencies used in our analysis, together with their uncertainties and the corresponding mass estimates, are listed in Table~\ref{table1}.

\begin{table*}[ht!]\centering
\begin{center}
\caption[]{\label{table1}Observed upper and lower twin-peak QPO frequencies of the selected microquasars and Galactic center source.}
\renewcommand{\arraystretch}{1.2}
\begin{tabular}{| l || c  c | c  c | l |}
\hline
Source 
&  $\nu_{\rm{U}}\,$[Hz]&$\Delta\nu_{\mathrm{U}}\,$[Hz]& $\nu_{\rm {L}}\,$[Hz]&$\Delta\nu_{\rm{L}}\,$[Hz]& Mass [\,M$_{\odot}$\,] \\
\hline
\hline
XTE~J1550-564 \cite{Orosz2011} & 276&$\pm\,3$& 184&$\pm\,5$&  ${9.1\pm 0.61}$\\
GRO~J1655--40 \cite{Strohmayer2001}    & 451&$\pm\,5$& 298&$\pm\,4$ & {5.4$\pm$0.3}   \\
GRS~1915+105 \cite{Miller2015}   & 168&$\pm\,3$& 113&$\pm\,5$&  ${12.4^{+2.0}_{-1.8}}$   \\
M82 X-1 \cite{Pasham2014}  & 5.07 & $\pm\,0.06$ & 3.32& $\pm\,0.06$ &$415\pm\,63$\\

\hline
        \end{tabular}
\end{center}
\end{table*}

As a first step, we perform a preliminary $\chi^2$ analysis to identify the region of the parameter space favored by the observed frequencies. Within the relativistic precession model, the theoretical QPO frequencies depend on four parameters, namely the black hole mass $M$, the magnetic charge parameter $q$, the PFDM parameter $\lambda$, and the orbital radius $r$ of the QPO-generating region. The corresponding $\chi^2$ statistic is written as \cite{Bambi2015GRO}
\begin{equation}
\chi^{2}(M,q,\lambda,r)=\frac{(\nu_{1\phi}-\nu_{1\rm U})^{2}}{\sigma_{1\rm U}^2}
+\frac{(\nu_{1\rm per}-\nu_{1\rm L})^{2}}{\sigma_{1 \rm L}^2}
+\frac{(\nu_{1\rm nod}-\nu_{1\rm C})^{2}}{\sigma_{1\rm C}^2}
+\frac{(\nu_{2\phi}-\nu_{2\rm U})^{2}}{\sigma_{2\rm U}^2}
+\frac{(\nu_{2\rm nod}-\nu_{2\rm C})^{2}}{\sigma_{2\rm C}^2}.
\end{equation}
The minimum of this function determines the most preferred parameter region, while the confidence intervals can be estimated from the vicinity of $\chi^2_{\min}$.

Using the outcome of this preliminary scan, we construct Gaussian priors for the parameters $M$, $q$, $\lambda$, and $r$. The corresponding mean values and standard deviations are summarized in Table~\ref{prior}. These priors are then used as inputs for the Bayesian inference, allowing the Markov chains to efficiently explore the physically relevant region without introducing unnecessarily restrictive assumptions.

\begin{table*}[ht!]\centering
\begin{center}
\renewcommand\arraystretch{1.5} 
\caption{\label{prior}Gaussian priors for $M$, $q$, $\lambda$, and $r$ for the selected sources.}
\begin{tabular}{lcccccccccc}
\hline\hline
\multirow{2}{*}{\textbf{RP}} & \multicolumn{2}{c}{XTE J1550-564} & \multicolumn{2}{c}{GRO J1655-40} & \multicolumn{2}{c}{GRS 1915+105} & \multicolumn{2}{c}{M82 X-1} \\
& $\mu$ & \multicolumn{1}{c}{$\sigma$} & $\mu$          & $\sigma$ & $\mu$ & \multicolumn{1}{c}{$\sigma$} & $\mu$ &\\
\hline
     $M/M_{\odot}$ & $8.8732$ & $1.0570$ & $5.3915$ & $0.6952$ & $14.7810$ & $1.8908$ & $480.485$ & $60.003$ &\\
    
     $q/M$ & $0.0585$ & $0.0361$   & $0.0598$  & $0.0362$ & $0.0622$ & $0.0377$ & $0.0648$ & $0.0403$  \\ 
    
     $\lambda/M$ & $0.2271$ & $0.1434$ & $0.2236$ & $0.1508$ & $0.2336$ & $0.1489$ & $0.2232$ & $0.1488$ \\
    
     $r/M$ & $5.0399$ & $0.5046$ & $5.0622$ & $0.5465$ & $4.988$ & $0.5278$ & $5.0538$ & $0.5130$ \\
     \hline\hline

\end{tabular}
\end{center}
\end{table*}

To derive the posterior distributions, we employ the Python package \texttt{emcee}, which implements the affine-invariant ensemble sampler for Markov Chain Monte Carlo (MCMC) calculations \cite{ForemanMackey2013}. The posterior probability density is defined through Bayes' theorem as \cite{Liu-etal2023,Hogg2018}
\begin{equation}
\mathcal{P}(\theta |\mathcal{D},\mathcal{M})=
\frac{P(\mathcal{D}|\theta,\mathcal{M})\,\pi(\theta|\mathcal{M})}
{P(\mathcal{D}|\mathcal{M})},
\end{equation}
where $\theta=\{M,q,\lambda,r\}$ denotes the parameter vector, $\pi(\theta|\mathcal{M})$ is the prior distribution, and $P(\mathcal{D}|\theta,\mathcal{M})$ is the likelihood function. Since the evidence term $P(\mathcal{D}|\mathcal{M})$ acts only as an overall normalization, the parameter estimation is effectively determined by the product of the prior and the likelihood.

In this analysis, the prior distributions are taken to be Gaussian within suitable parameter bounds,
\begin{equation}
\pi(\theta_i)\propto
\exp\left[
-\frac{1}{2}
\left(
\frac{\theta_i-\theta_{0,i}}{\sigma_i}
\right)^2
\right],
\qquad
\theta_{{\rm low},i}<\theta_i<\theta_{{\rm high},i},
\end{equation}
where $\theta_{0,i}$ and $\sigma_i$ denote the mean values and standard deviations inferred from the preliminary $\chi^2$ analysis. The log-likelihood function is expressed as the sum of the contributions from the upper and lower QPO frequencies,
\begin{equation}
\log \mathcal{L} = \log \mathcal{L}_{\rm U} + \log \mathcal{L}_{\rm L},
\label{likelyhood}
\end{equation}
where
\begin{equation}
\log \mathcal{L}_{\rm U}
=
-\frac{1}{2}\sum_i
\left(
\frac{\nu^i_{{\rm U,obs}}-\nu^i_{{\rm U,th}}}
{\sigma^i_{{\rm U,obs}}}
\right)^2,
\end{equation}
and
\begin{equation}
\log \mathcal{L}_{\rm L}
=
-\frac{1}{2}\sum_i
\left(
\frac{\nu^i_{{\rm L,obs}}-\nu^i_{{\rm L,th}}}
{\sigma^i_{{\rm L,obs}}}
\right)^2.
\end{equation}
Here, $\nu^i_{{\rm U,obs}}$ and $\nu^i_{{\rm L,obs}}$ are the observed upper and lower QPO frequencies, while $\nu^i_{{\rm U,th}}$ and $\nu^i_{{\rm L,th}}$ are the corresponding theoretical predictions of the model. The quantities $\sigma^i_{{\rm U,obs}}$ and $\sigma^i_{{\rm L,obs}}$ denote the observational uncertainties.

\begin{figure}[tbhp]
    \centering
    \includegraphics[width=0.49\linewidth]{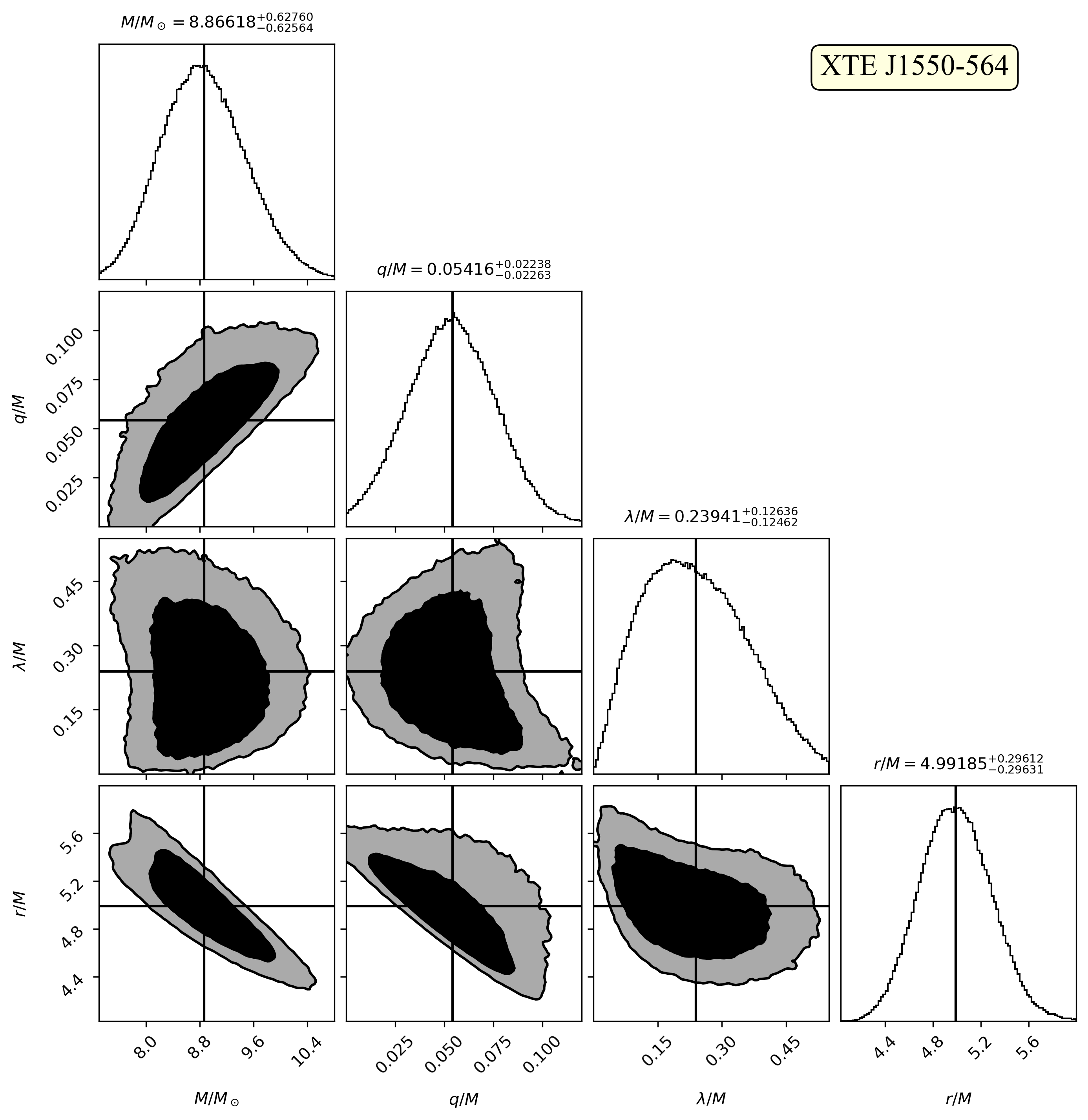}
    \includegraphics[width=0.49\linewidth]{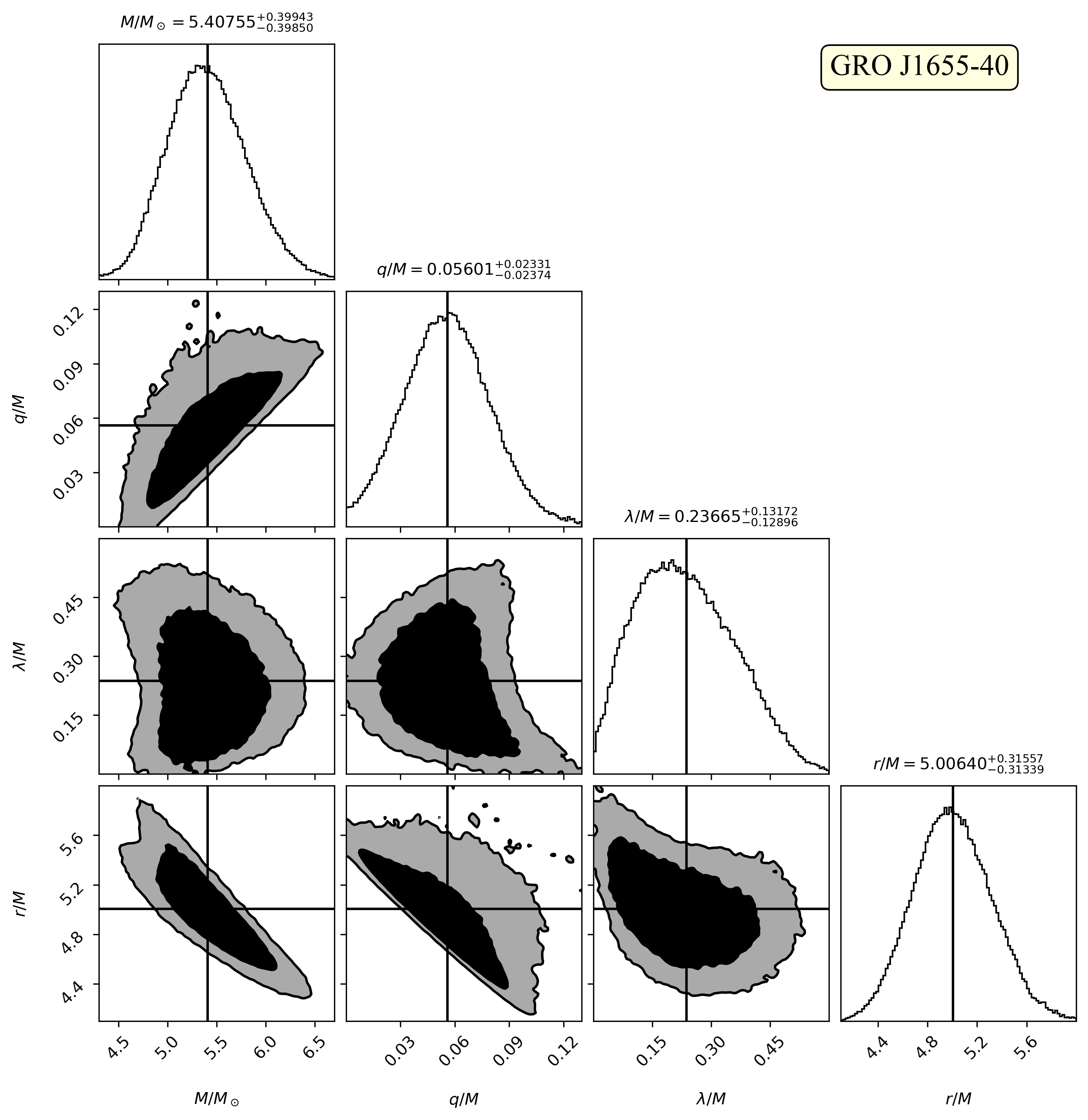}
    \includegraphics[width=0.49\linewidth]{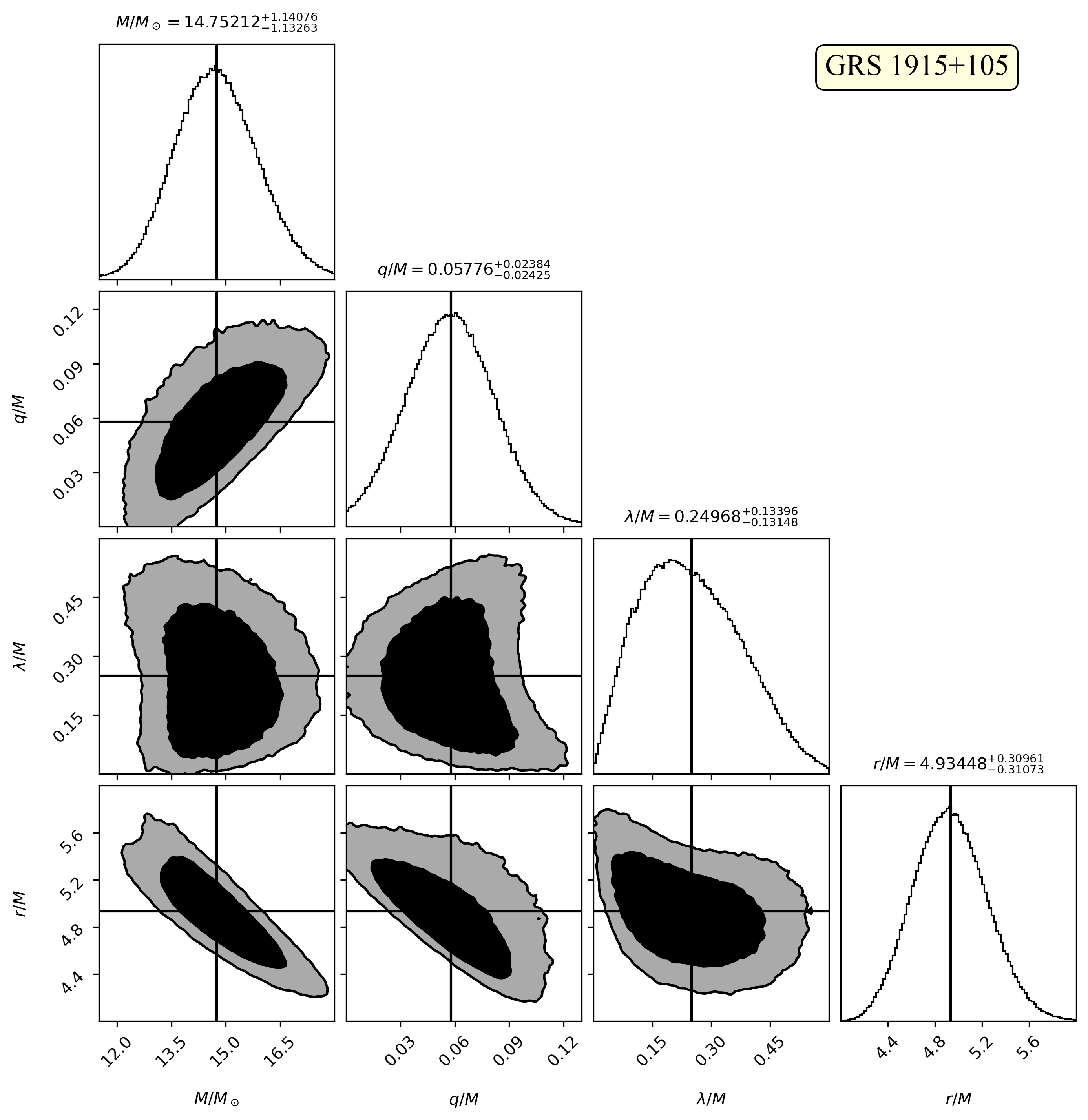}
    \includegraphics[width=0.49\linewidth]{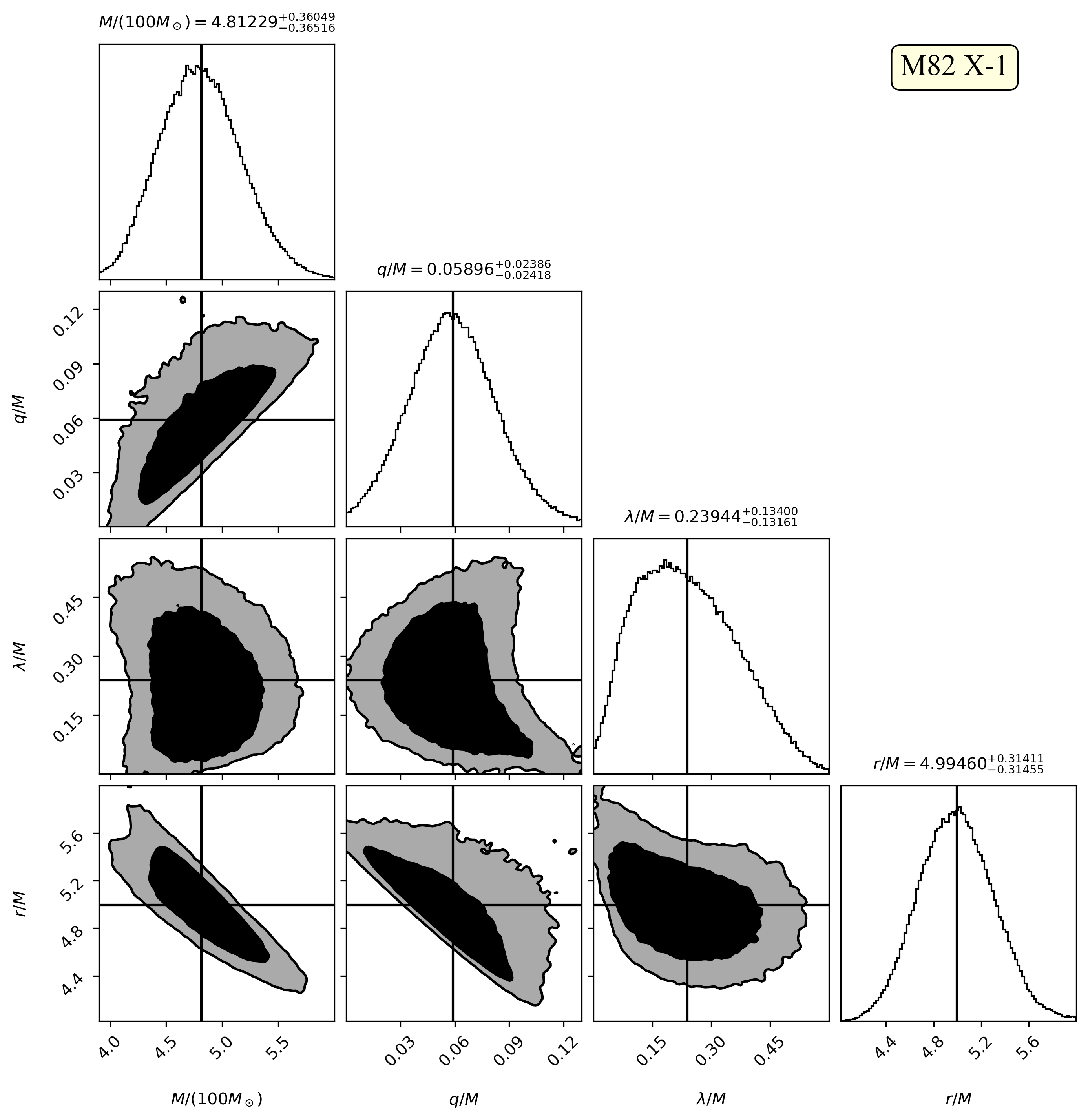}
    \caption{Posterior distributions of $M$, $q$, $\lambda$, and $r$ obtained from the MCMC analysis in the RP model for XTE~J1550--564, GRO~J1655--40, GRS~1915+105, and M82~X-1.\label{XTE}}
    \label{fig:1b}
\end{figure}

Next, we run the MCMC chains to constrain the four parameters $(M,q,\lambda,r)$ of the regular black hole immersed in PFDM. In practice, we generate approximately $5\times10^4$ samples for each parameter, which allows a broad exploration of the admissible parameter space and yields stable posterior estimates. The resulting one-dimensional marginalized distributions and two-dimensional confidence regions are displayed in Fig.~\ref{XTE}.

Figure~\ref{XTE} presents the posterior distributions obtained from the MCMC analysis for the four selected sources. The diagonal panels show that all parameters possess well-defined marginalized distributions with clear maxima, which indicates that the observational QPO data are capable of placing meaningful constraints on the model. In particular, the black hole mass is tightly constrained in each case and remains consistent with the known astrophysical expectations for the corresponding source. The distributions of the magnetic charge parameter $q$ are centered around small positive values, suggesting that only modest deviations from the Schwarzschild case are favored by the data. A similar behavior is found for the PFDM parameter $\lambda$, whose posterior remains restricted to a narrow positive interval. Another important result is that the orbital radius associated with the QPO generation region is strongly localized around $r/M \approx 5$ for all sources, which supports the interpretation that the observed oscillations originate in the inner part of the accretion flow, close to the strong-gravity region.

The two-dimensional contours provide additional information about correlations between the parameters. A pronounced negative correlation is visible between $M$ and $r$, indicating that a larger black hole mass can be compensated by a smaller orbital radius in order to reproduce the same pair of QPO frequencies. The plots also reveal a positive correlation between $M$ and $q$, which implies that an increase in the magnetic charge parameter tends to shift the preferred mass toward larger values. By contrast, the contours involving $\lambda$ are broader and less tilted, showing that the PFDM parameter is somewhat less correlated with the remaining quantities, although it still influences the allowed parameter region. Overall, the similarity of the posterior structure across XTE~J1550--564, GRO~J1655--40, GRS~1915+105, and M82~X-1 suggests that the model behaves consistently over a wide mass range, from stellar-mass systems to an intermediate-mass black hole candidate.

The posteriors show that the model provides well-defined constraints for all four sources. In particular, the black hole mass is tightly bounded in each case and remains consistent with the expected astrophysical range of the corresponding object. The magnetic charge parameter $q$ is systematically restricted to small positive values, while the PFDM parameter $\lambda$ is confined to a relatively narrow interval. Another robust feature is that the preferred orbital radius is localized around $r/M\sim5$, indicating that the observed QPOs originate in the inner region of the accretion flow, close to the strong-field zone of the spacetime. The final best-fit values and associated uncertainties are reported in Table~\ref{tab3}.

\begin{table*}\centering
\centering
\renewcommand{\arraystretch}{1.5}
\setlength{\tabcolsep}{8pt}
\begin{tabular}{|l|p{3cm}|p{2cm}|p{2cm}|p{2cm}|p{2cm}|}
\hline
\textbf{Model: RP} & ${M/(M_\odot)}$  & $q/M$ & $\lambda/M$ & $r/M$ \\
\hline
XTE~J1550-564 & $8.8662^{+0.6276}_{-0.6256}$ & \(0.0542^{+0.0224}_{-0.0226}\)  & \(0.2394^{+0.1264}_{-0.1246}\) &\(4.9918^{+0.2961}_{-0.2963}\) \\

\hline
GRO~J1655--40 & \(5.4075^{+0.3994}_{-0.3985}\) & \(0.0560^{+0.0233}_{-0.0237}\) & \(0.2366^{+0.1317}_{-0.1290}\) &\(5.0064^{+0.3156}_{-0.3134}\)\\

\hline
GRS~1915+105 & \(14.7521^{+1.1408}_{-1.1326}\) & \(0.0578^{+0.0238}_{-0.0242}\) & \(0.2497^{+0.1340}_{-0.1315}\) &\(4.9345^{+0.3096}_{-0.3107}\)\\

\hline
M82 X-1 & \( 481.229^{+36.049}_{-36.516}\) & \(0.0590^{+0.0239}_{-0.0242}\) & \(0.2394^{+0.1340}_{-0.1316}\) &\(4.9946^{+0.3141}_{-0.3145}\)\\
\hline
\end{tabular}
\caption{Best-fit values of the RP model parameters for the selected black hole sources.}
\label{tab3}
\end{table*}

Overall, the MCMC analysis demonstrates that the relativistic precession model applied to a regular black hole surrounded by perfect fluid dark matter can successfully reproduce the observed twin-peak QPO data. At the same time, it provides simultaneous constraints on the black hole mass and on the additional spacetime parameters introduced by the magnetic charge and the PFDM environment.

\section{Geodesic motion and black hole shadow}\label{Sec:shadow}

The analysis of test particle trajectories in the vicinity of a black hole plays a central role in determining the observable shape and size of the black hole shadow. Let us consider a test particle of rest mass $m_0$ moving in this spacetime. The derivation of geodesic equations becomes more transparent once the conserved quantities associated with the spacetime symmetries are identified. Denoting by $y^\sigma$ the vector corresponding to a symmetry direction and by $v^\sigma = dx^\sigma/d\lambda$ the tangent vector to the particle worldline $x^\sigma=x^\sigma(\tau)$, where $\tau$ is an affine parameter, one finds that for geodesic motion the contraction of the Killing vector with the four-velocity remains constant \cite{raine2010black}, namely

\begin{equation}\label{e6}
y^{\sigma}v_{\sigma}=constant.
\end{equation}

For a static spacetime, the Killing vector associated with time translations is $y^\sigma=(1,0,0,0)$, whereas axial symmetry gives the azimuthal Killing vector $y^\sigma=(0,0,0,1)$. Using Eq.~(\ref{e6}), the corresponding conserved quantities can be written as

\begin{equation}\label{e7}
y^{0}v^{0}=v^{0}=-\mathrm{E},\qquad y^{3}v_{3}=v_{3}=\mathrm{L}.
\end{equation}

Here, $\mathrm{E}$ and $\mathrm{L}$ denote, respectively, the conserved energy and angular momentum per unit mass of the particle. These constants immediately yield two of the geodesic equations in the form

\begin{eqnarray}\label{e9}
&&u^{0}=g^{0\mu}u_{\mu}=g^{00}u_{0}=\frac{\mathrm{E}}{f(r)},\\
&&u^{3}=g^{3\mu}u_{\mu}=g^{33}u_{3}=\frac{\mathrm{L}}{r^{2} \sin^{2}\theta}.
\end{eqnarray}

or equivalently,

\begin{equation}\label{e10}
    \frac{dt}{d\tau}=\frac{\mathrm{E}}{f(r)}, \qquad \frac{d\phi}{d\tau}=\frac{\mathrm{L}}{r^{2} \sin^{2}\theta}.
\end{equation}

To obtain the remaining equations of motion, we employ the Hamilton--Jacobi formalism. The corresponding Hamilton--Jacobi equation is

\begin{eqnarray}\label{e11}
\frac{\partial S}{\partial \tau}+\frac{1}{2}g^{\mu\sigma}\frac{\partial S}{\partial x^{\mu}}\frac{\partial S}{\partial x^{\sigma}}=0.
\end{eqnarray}

Following the standard separation procedure outlined in Ref.~\cite{Chandrasekhar1983}, the action may be written as

\begin{equation}\label{e12}
H=H_{r}-\mathrm{E} t+\mathrm{L}\phi+H_{\theta}+\frac{1}{2}m_{0}^{2}\tau.
\end{equation}

In this expression, $H_r$ and $H_\theta$ depend only on $r$ and $\theta$, respectively. Substituting the separated action into Eq.~(\ref{e11}) leads to the radial and angular parts

\begin{eqnarray}\label{e13}
H_{r}=\int^{r}\frac{\sqrt{X_r}}{r^{2}f(r)}dr, \qquad
H_{\theta}=\int^{r}\sqrt{X_\theta}d\theta,
\end{eqnarray}

where

\begin{eqnarray}\label{e14}
X_r&=&r^{4}\mathrm{E}^{2}-r^{2}\left(r^{2}m_{0}^{2}+K+\mathrm{L}^{2}\right)f(r),\\
X_\theta&=& K-\mathrm{L}^{2}\cot^{2}\theta.
\end{eqnarray}

Thus, for a particle moving in a static black hole background, the geodesic equations corresponding to the polar and radial directions become

\begin{equation}\label{e151}
r^{2}\frac{d\theta}{d\tau}=\sqrt{K-\mathrm{L}^{2}\cot^{2}\theta},
\end{equation}
\begin{equation}\label{e161}
r^{2}\frac{d r}{d\tau}=\sqrt{r^{4}\mathrm{E}^{2}-r^{2}\left(r^{2}m_{0}^{2}+K+\mathrm{L}^{2}\right)f(r)}.
\end{equation}

Equation~(\ref{e151}) governs the motion in the $\theta$ direction, while Eq.~(\ref{e161}) determines the radial evolution. For photon trajectories, one sets $m_0^2=0$, and the null geodesic equations near the black hole reduce to \cite{a26}

\begin{equation}\label{e17}
r^{2}\frac{d\theta}{d\tau}=\sqrt{K-\mathrm{L}^{2}\cot^{2}\theta},
\end{equation}
\begin{equation}\label{e18}
r^{2}\frac{d r}{d\tau}=\sqrt{r^{4}\mathrm{E}^{2}-r^2\left(K+\mathrm{L}^{2}\right)f(r)}.
\end{equation}

The radial equation may be recast in the form

\begin{equation}\label{e19}
\left(\frac{d r}{d \tau}\right)^{2}+V_{\rm eff}=0,
\end{equation}

where the effective potential is defined as

\begin{equation}\label{e20}
V_{\rm eff}=\frac{f(r)}{r^{2}}\left(K+\mathrm{L}^{2}\right)-\mathrm{E}^{2}.
\end{equation}

The unstable circular photon orbit, which is crucial for shadow formation, is determined from the standard conditions

\begin{eqnarray}\label{e21}
V_{\rm eff}(r=r_{p})=V^{'}_{\rm eff}(r=r_{p})=0,
\end{eqnarray}

with $r_p$ representing the photon sphere radius and the prime denoting differentiation with respect to $r$. The first condition, $V_{\rm eff}(r_p)=0$, yields

\begin{equation}\label{ee22}
\frac{r_{p}^{2}}{f(r_{p})}=\xi+\eta^{2}.
\end{equation}

Here, the Chandrasekhar constants \cite{Chandrasekhar1983} are introduced as $\xi={K}/{\mathrm{E}^{2}}$ and $\eta=\mathrm{L}/\mathrm{E}$. The second condition, $V'_{\rm eff}(r_p)=0$, gives

\begin{eqnarray}\label{e24}
r_{p}f^{'}(r_{p})-2f(r_{p})=0 \Longrightarrow 1 - \frac{\lambda}{2r_p} - \frac{3 M r^3_p}{(r_p+q)^4} + \frac{3 \lambda}{2r_p} \ln\!\frac{r_p}{\lambda}=0.\label{photon-radius}
\end{eqnarray}

\begin{table}[ht!]
\centering
\begin{tabular}{|c|ccccc|}
\hline
$\lambda/M (\downarrow) \backslash q/M (\rightarrow)$ & 0.1 & 0.2 & 0.3 & 0.4 & 0.5 \\
\hline
-0.5 & 4.0367 & 3.6720 & 3.2703 & 2.8015 & 2.1171 \\
-0.4 & 3.8709 & 3.5052 & 3.1033 & 2.6365 & 1.9727 \\
-0.3 & 3.6713 & 3.2999 & 2.8905 & 2.4118 & 1.7011 \\
-0.2 & 3.4262 & 3.0420 & 2.6135 & 2.0976 & 1.1904 \\
-0.1 & 3.1087 & 2.6979 & 2.2235 & 1.5815 & 1.1166 \\
\hline
\end{tabular}
\caption{Photon sphere radius $r_{\text{ph}}/M$ for different values of $q/M$ and $\lambda/M$.}
\label{tab:4}
\end{table}

The exact analytical solution of the above polynomial gives the photon sphere radius. Noted that the exact solution is impossible due to the presence of the logarithmic function. Thereby, choosing suitable values of the parameters $\lambda$ and $q$, one can determine the photon sphere radius. In Table \ref{tab:4}, we presented numerical values of the photon sphere radius $r_p$ by varying $\lambda$ and $q$.

Now consider photons emitted by a luminous source located behind the black hole relative to the observer. Depending on their initial conditions, these photons may either escape to infinity after being deflected, be captured by the black hole, or move along a critical orbit separating these two outcomes. The photons that reach the observer contribute to the visible image, whereas those captured by the black hole produce a dark region in the observer’s sky. This dark area is identified as the black hole shadow. In this section, our aim is to explore the shadow structure generated by the black hole in PFDM spacetime. For this purpose, we first introduce the celestial coordinates \cite{papnoi2014shadow,Zahid2021ChJPh..72..575Z,Zahid:2024nvx},

\begin{eqnarray}\label{eq28}
\alpha&=&\lim_{r \to \infty}-\left(r^{2} \sin\theta\frac{d\phi}{dr}\right),\\
\beta&=&\lim_{r \to \infty}\left(r^{2}\frac{d\theta}{dr}\right).
\end{eqnarray}

In these expressions, $\alpha$ measures the apparent horizontal displacement of the shadow with respect to the symmetry axis, while $\beta$ gives its apparent vertical position in the observer’s sky. The angle $\theta$ is the inclination between the observer’s line of sight and the symmetry axis, and $r_0$ denotes the observer’s distance from the black hole. Using the geodesic equations, the derivatives entering Eq.~(\ref{eq28}) can be expressed as

\begin{equation}\label{eq29}
\frac{d\phi}{dr}=\frac{\mathrm{L} \csc^{2}\theta}{r^{2}\sqrt{\mathrm{E}^{2}-\frac{f(r)}{r^{2}}\left(K+\mathrm{L}^{2}\right)}},
\end{equation}
\begin{equation}\label{eq30}
\frac{d\theta}{dr}=\frac{1}{r^{2}}\sqrt{\frac{K - \mathrm{L}^{2}\cot^{2}\theta}{\mathrm{E}^{2}-\frac{f(r)}{r^{2}}\left(K+\mathrm{L}^{2}\right)}}.
\end{equation}

Taking the limit $r\to\infty$, one obtains the celestial coordinates in the form \cite{Zahid:2024hyy,zahid2024shadow}

\begin{equation}\label{e31}
\alpha=-\eta\csc^2(\theta ) \sin (\theta ),\quad \beta=\pm\sqrt{\xi-\eta^2\cot^2(\theta)}.
\end{equation}

For an observer located in the equatorial plane, $\theta=\pi/2$, these expressions simplify to

\begin{equation}\label{e33}
\alpha=-\eta,   \qquad    \beta=\pm\sqrt{\xi}.
\end{equation}

Accordingly, the shadow radius in the celestial plane $(\alpha,\beta)$ is

\begin{equation}\label{e35}
R^{2}_{s}=\alpha^{2}+\beta^{2}=\eta^2+\xi=\frac{r_{p}^{2}}{f(r_{p})}.
\end{equation}

Here, $R_s$ denotes the shadow radius for the non-rotating black hole.

\begin{figure}
    \centering
    \includegraphics[width=0.95\linewidth]{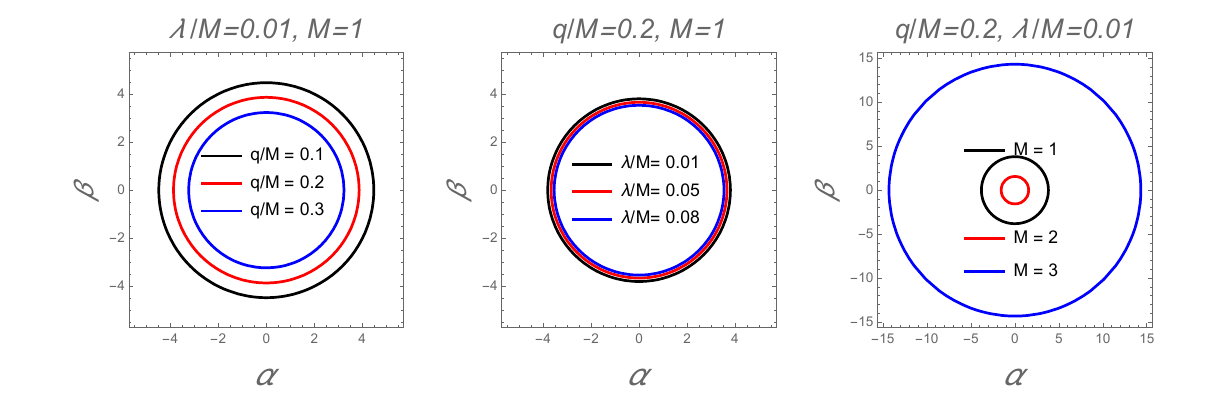}
   \caption{Shadow profiles in the celestial plane $(\alpha,\beta)$ for different values of $q/M$, $\lambda/M$, and $M$. The left panel shows the dependence on the magnetic charge parameter $q/M$, the middle panel illustrates the effect of the PFDM parameter $\lambda/M$, and the right panel presents the variation with the black hole mass $M$.}
    \label{shadoww}
\end{figure}

Figure \ref{shadoww} shows the black hole shadow in the celestial coordinates $(\alpha,\beta)$ for different values of $q/M$, $\lambda/M$, and $M$. In general, the shadow remains nearly circular, while its overall size changes noticeably with the parameters. This indicates that the magnetic charge, the PFDM contribution, and the black hole mass all affect the photon capture region and hence the apparent shadow.
In the left panel, the shadow is shown for different values of $q/M$ with fixed $\lambda/M=0.01$ and $M=1$. It is clear that the shadow radius decreases as $q/M$ increases. This behavior implies that a stronger magnetic charge parameter reduces the size of the photon capture region, leading to a smaller apparent shadow.
In the middle panel, the shadow is plotted for different values of $\lambda/M$ with fixed $q/M=0.2$ and $M=1$. One can see that increasing $\lambda/M$ also decreases the shadow size, although the effect is comparatively weaker than in the case of $q/M$. This shows that the PFDM environment modifies the propagation of photons and slightly shrinks the shadow boundary.
In the right panel, the shadow is presented for different values of $M$ with fixed $q/M=0.2$ and $\lambda/M=0.01$. In this case, the shadow size grows significantly with increasing mass. This is physically expected, since a larger black hole produces a wider photon capture region and therefore a larger shadow in the observer's sky.

\begin{figure}[ht!]
    \centering
    \includegraphics[width=0.45\linewidth]{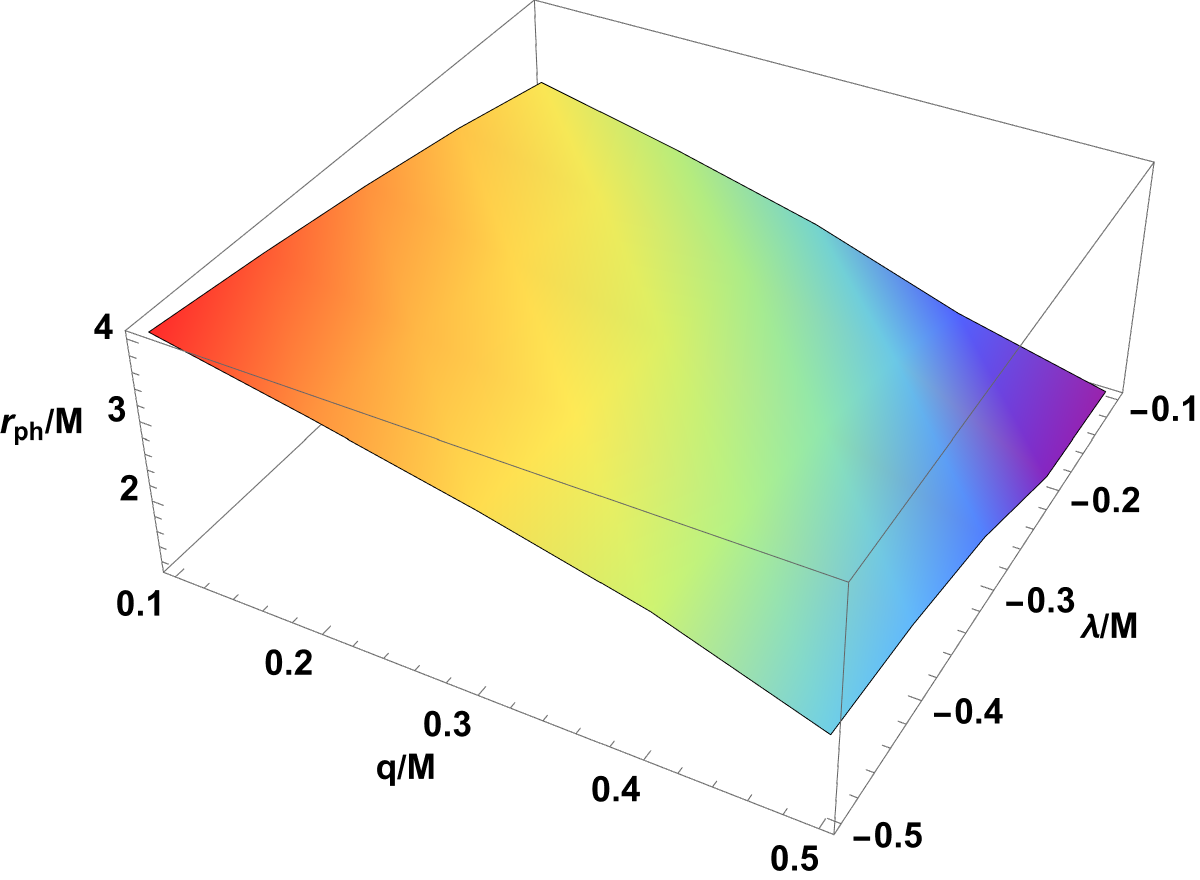}\qquad
    \includegraphics[width=0.45\linewidth]{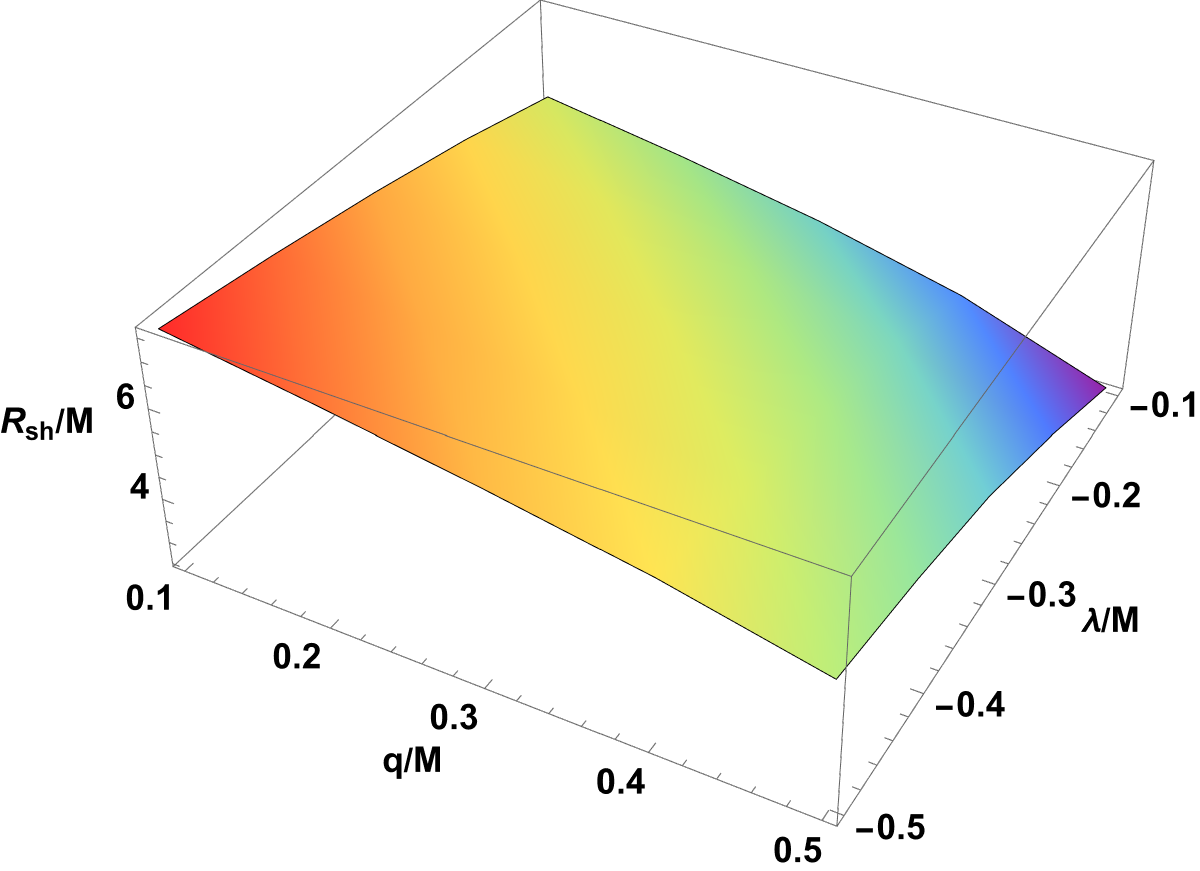}
    \caption{Three-dimensional behavior of the photon sphere radius and the shadow radius for different values of $q$ and $\lambda$.}
    \label{fig:3}
\end{figure}

Figure \ref{fig:3} shows the three-dimensional behavior of the photon-sphere radius $r_{\rm ph}/M$ and the shadow radius $R_{\rm sh}/M$ as functions of the magnetic charge $q/M$ and the PFDM parameter $\lambda/M$. In both panels, the surfaces decrease monotonically, indicating that increasing $q/M$ and shifting $\lambda/M$ toward larger values in the plotted range reduce both the photon-sphere radius and the shadow size. This behavior implies that the combined effect of nonlinear electrodynamics and the perfect fluid dark matter environment moves the unstable circular photon orbit closer to the black hole, thereby shrinking the photon capture region. As a result, the apparent black hole shadow becomes smaller. It is also seen that the dependence on the magnetic charge is more pronounced than that on the PFDM parameter, which suggests that the nonlinear-electrodynamics sector provides the leading correction to the null-geodesic structure, while PFDM acts as an additional environmental modification. These results show that both parameters leave a direct imprint on the optical appearance of the black hole and may therefore be relevant for distinguishing this spacetime from the Schwarzschild case through shadow observations.

\begin{table}[ht!]
\centering
\begin{tabular}{|c|ccccc|}
\hline
$\lambda/M (\downarrow) \backslash q/M (\rightarrow)$ & 0.1 & 0.2 & 0.3 & 0.4 & 0.5 \\
\hline
-0.5 & 7.6162 & 7.1472 & 6.6359 & 6.0549 & 5.3083 \\
-0.4 & 7.2276 & 6.7555 & 6.2404 & 5.6552 & 4.9066 \\
-0.3 & 6.7851 & 6.3041 & 5.7768 & 5.1724 & 4.3738 \\
-0.2 & 6.2690 & 5.7705 & 5.2171 & 4.5654 & 3.5323 \\
-0.1 & 5.6346 & 5.1014 & 4.4904 & 3.7024 & 2.5372 \\
\hline
\end{tabular}
\caption{Shadow radius $R_{\text{s}}/M$ as a function of $q/M$ and $\lambda/M$.}
\label{tab:5}
\end{table}

In Table \ref{tab:5}, we presented numerical values of the shadow radius $R_{\rm s}$ by varying $\lambda$ and $q$. Moreover, In Figure \ref{fig:3}, we present graphical representations of the photon sphere radius and shadow radius as a function of the PFDM parameter $\lambda$ and the magnetic charge $q$.

\section{Conclusion}\label{sec8}

In this work, we have studied a black hole solution supported by nonlinear electrodynamics and embedded in a perfect fluid dark matter environment, focusing on its thermodynamic behavior, test-particle dynamics, QPO phenomenology, and shadow properties. The results show that the magnetic charge parameter $q$ and the PFDM parameter $\lambda$ produce clear departures from the Schwarzschild case and leave observable imprints on both the dynamical and optical properties of the spacetime.

The thermodynamic analysis shows that the horizon structure is significantly modified by the model parameters. In particular, the Hawking temperature no longer follows the standard monotonic Schwarzschild behavior, but instead develops a non-monotonic profile with a finite maximum. At the same time, the heat capacity exhibits divergences and sign changes, indicating second-order phase transitions and the emergence of a locally stable small-horizon branch. These features demonstrate that the combined NED--PFDM background gives rise to a much richer thermal structure than in the vacuum case.

The study of neutral test-particle motion further shows that the effective potential, circular orbit characteristics, and epicyclic frequencies are all sensitive to the spacetime deformation. The orbital energy and angular momentum differ noticeably from the Schwarzschild case, while the numerical ISCO analysis shows that the innermost stable circular orbit decreases in the considered parameter range as $q$ and $\lambda$ increase. The radial epicyclic frequency and the associated relativistic-precession relation are also shifted, implying that the QPO spectrum carries direct information about both the nonlinear-electrodynamics sector and the surrounding dark matter environment.

Using the observed twin-peak QPO data of XTE J1550--564, GRO J1655--40, GRS 1915+105, and M82 X-1, we performed an MCMC analysis and obtained meaningful constraints on the model parameters. The posteriors consistently favor small positive values of the magnetic charge, a relatively narrow interval for the PFDM parameter, and a preferred QPO generation region around $r/M \sim 5$. This indicates that the model is capable of reproducing the observed timing data across a broad mass range, from stellar-mass black hole candidates to an intermediate-mass source.

Finally, the null-geodesic analysis shows that both the photon-sphere radius and the shadow radius decrease with increasing $q$ and $\lambda$ in the considered range. This means that the combined effect of nonlinear electrodynamics and perfect fluid dark matter shifts the unstable photon orbit inward and reduces the photon capture region, leading to a smaller apparent shadow. The dependence on the magnetic charge is more pronounced, indicating that the NED sector provides the dominant correction to the optical structure, while PFDM acts as an additional environmental contribution.

These results show that black holes in nonlinear electrodynamics surrounded by perfect fluid dark matter exhibit nontrivial thermodynamic phases, modified orbital dynamics, measurable changes in QPO relations, and distinct shadow signatures. This makes the model a useful framework for connecting strong-field gravity with astrophysical timing and imaging observations, and for probing possible signatures of nonlinear electromagnetic effects and dark matter around compact objects.

\scriptsize

\section*{Acknowledgments}

F.A. acknowledges the Inter University Centre for Astronomy and Astrophysics (IUCAA), Pune, India for granting visiting associateship.\\ S.M. gratefully acknowledges support from Grant FZ-20200929385 of the Agency of Innovative Developments of the Republic of Uzbekistan.

\bibliographystyle{apsrev4-1}

\bibliography{sample}

\end{document}